\documentclass[%
 reprint,
superscriptaddress,
 amsmath,amssymb,
 aps,
]{revtex4-2}

\usepackage{graphicx}
\usepackage{dcolumn}
\usepackage{bm}
\usepackage{xcolor}
\usepackage{soul}

\usepackage{amsmath}
\usepackage{braket}
\usepackage{siunitx}   
\usepackage{braket}

\def\orcid#1{\kern .08em\href{https://orcid.org/#1}{\includegraphics[keepaspectratio,width=0.7em]{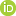}}}

\begin{document}

\preprint{APS/123-QED}

\title[Demonstration of a Raman Velocity  Filter in Collinear Laser Spectroscopy]{Demonstration of a Raman Velocity  Filter in Collinear Laser Spectroscopy: Towards Applications for sub-ppm High-Voltage Measurements} 
\author{J. Spahn\orcid{0009-0007-8354-4896}}
    \affiliation{Institut für Kernphysik, Technische Universität Darmstadt, 64289 Darmstadt, Germany}
    \email{jspahn@ikp.tu-darstadt.de}
\author{H. Bodnar}
    \affiliation{Institut für Kernphysik, Technische Universität Darmstadt, 64289 Darmstadt, Germany}
\author{Kristian König\orcid{0000-0001-9415-3208}}
    \affiliation{Institut für Kernphysik, Technische Universität Darmstadt, 64289 Darmstadt, Germany}%
    \affiliation{Helmholtz Research Academy Hesse for FAIR, GSI Helmholtzzentrum für Schwerionenforschung, 64291 Darmstadt, Germany}
\author{W. Nörtershäuser\orcid{0000-0001-7432-3687}}
    \affiliation{Institut für Kernphysik, Technische Universität Darmstadt, 64289 Darmstadt, Germany}%
    \affiliation{Helmholtz Research Academy Hesse for FAIR, GSI Helmholtzzentrum für Schwerionenforschung, 64291 Darmstadt, Germany}


\date{\today}

\newcommand{\Sone}{$\mathrm{S}_{1/2}$}
\newcommand{\Pthree}{$\mathrm{P}_{3/2}$}
\newcommand{\Dthree}{$\mathrm{D}_{3/2}$}
\newcommand{\Dfive}{$\mathrm{D}_{5/2}$}

\newcommand{\SonePthree}{$\mathrm{S}_{1/2}\rightarrow\mathrm{P}_{3/2}$}
\newcommand{\DthreePthree}{$\mathrm{D}_{3/2}\rightarrow\mathrm{P}_{3/2}$}
\newcommand{\DfivePthree}{$\mathrm{D}_{5/2}\rightarrow\mathrm{P}_{3/2}$}

\newcommand{\SoneDthree}{$\mathrm{S}_{1/2}\rightarrow\mathrm{D}_{3/2}$}
\newcommand{\SoneDfive}{$\mathrm{S}_{1/2}\rightarrow\mathrm{D}_{5/2}$}
\newcommand{\DthreeDfive}{$\mathrm{D}_{3/2}\rightarrow\mathrm{D}_{5/2}$}

\begin{abstract}
Raman transitions have a wide range of applications in atomic physics and have recently been proposed as a means for improving high-precision high-voltage measurements. Here, we present a theoretical analysis and a first experimental demonstration of $5s\,^2\mathrm{S}_{1/2} \rightarrow 4d\,^2\mathrm{D}_{3/2,5/2}$ Raman transitions in $^{88}$Sr$^+$ ions in collinear laser spectroscopy. For the theoretical description the three-level system is reduced to an effective two-level system, in order to estimate the experimental parameters, while the role of the spatial laser intensity distribution in combination with the radial extension of the ion beam are elucidated by performing simulations of the full four-level system. Experimentally, we realized the first velocity-selective Raman transition in collinear laser spectroscopy. Using a $^{88}$Sr$^+$ ion beam, we demonstrate a reduction in the energy width to less than $200\,$meV, which is about an order of magnitude reduction compared to the usage of an optical dipole transition as in previous works. We also investigate two-photon Rabi oscillations and show that their observed collapse is consistent with the simulations.

\end{abstract}

\maketitle


\section{\label{sec:Introduction}Introduction}

\subsection{\label{ssec:Motivation}Motivation}

The interest in precise definitions and measurements of units in physics as well as industry has led to considerable work and effort being invested in the field of metrology over the last two centuries. The first steps toward an international uniform and coherent system of units were taken in 1875 with the signing of the Metre Convention, ultimately leading to the birth of the \textit{System International} (SI)  in 1960 \cite{gupta2020metre}. Since then, many of the original definitions have been revised, aiming to ascribe the definition and measurement of units to fundamental constants and quantum effects.\\ 
For example, the experimental observation of the Josephson effect in 1963 \cite{josephson1962possible, anderson1963probable} opened up the possibility of measuring voltages through this quantum effect in Josephson junctions, two superconductors separated by a thin isolator, and radio frequency measurements. In 1990, the first quantum electrical standard was launched and inconsistencies with the SI definitions of the electrical units were resolved with the revision of the SI in 2019 \cite{stock2019revision}. 
Connecting several tens of thousands of Josephson junctions in series on dedicated chips allows measuring voltages up to $10\,$V with down to $2\cdot10^{-11}$ relative uncertainty \cite{rufenacht2018impact}. As measuring higher voltages requires stacking more Josephson junctions, increasing the complexity of the chip design, high-voltage measurements using this technique are not feasible. Instead, high-voltage dividers are used to scale down the high voltage to a range where it can be referenced to Josephson voltage standards. This requires precise knowledge of the divider ratio which depends on the resistance of the resistor chain of the divider. The latter varies with ambient conditions and time, limiting this method to ppm accuracy \cite{MarxBPTHVDiv, thummler2009precision, passon20243DHVDiv}.\\
In 1982, Poulsen proposed an alternative approach. The Doppler-shifted atomic transition frequency of ions, that have been electrostatically accelerated using the high voltage of interest, is measured by collinear laser spectroscopy \cite{poulsen1982velocity}. The high voltage is then extracted from the measured Doppler shift. 
First experiments exploring this technique were performed by Poulsen and Riis \cite{OPoulsen_1988}, who proposed selecting a specific velocity class via optical pumping, and later experiments trying to develop this technique towards ppm accuracy were performed by Götte \textit{et al.}~\cite{gotte2004test}. 
Latest measurements on $^{40}$Ca$^+$ beams at the Technical University of Darmstadt using this technique have matched the parts-per-million (ppm) relative uncertainty of high-voltage dividers \cite{kramer2018high}.\\
In this ongoing effort to reach sub-ppm accuracy, the energy width of the ion beam is still a potentially limiting factor. This could be overcome by using a Raman transition to prepare ions in a metastable state for subsequent high-precision measurements. As the intrinsic linewidth of a Raman transition is orders of magnitude smaller than the natural linewidth of a dipole transition, this would serve as an excellent velocity filter.
While using even narrower but slow clock transitions is not feasible in collinear laser spectroscopy due to the short interaction times of typically a few $\mathrm{\mu s}$, detailed calculations indicating the feasibility of this approach in Ca$^+$ have recently been performed by Neumann \textit{et al.}~\cite{neumann2020raman}. 
However, an experimental proof-of-concept is still missing, as driving a stimulated Raman transition in collinear laser spectroscopy has so far only been demonstrated using a single laser and its RF-shifted sideband \cite{dinneen1991stimulated}, but not using two separate lasers, as required for the optical high-voltage measurements.\\
This challenge was addressed in the $\mathrm{S}_{1/2}$, $\mathrm{P}_{3/2}$, $\mathrm{D}_\mathrm{J}$ ($\mathrm{J}\in\{3/2,5/2\}$) $\Lambda$-scheme of $^{88}$Sr$^+$, which has a similar level scheme as $^{40}$Ca$^+$. Experimental limitations were explored, and the dynamics of two-photon Rabi oscillations were investigated and compared to simulations that include spatial intensity distributions of the laser and ion beams.\\

\subsection{\label{ssec:CLS}Collinear Laser Spectroscopy}
Collinear laser spectroscopy (CLS)\cite{anton1978collinear} is a fast technique to precisely investigate atomic transition frequencies, and hence, widely used for measurements in short-lived isotopes \cite{Schinzler1978, otten1989shortlived, yang2023laser} that provide access to nuclear properties such as the charge radius or electromagnetic moments. Various CLS experiments have been established at a wide range of online facilities 
to investigate isotopes across the entire nuclear landscape \cite{Blaum_2013rev, Campbell_2016rev, yang2023laser}.
\\  
In CLS, laser spectroscopy is performed on an ion beam or neutral atom beam by superimposing the beam with a collinear (copropagating) and/or anticollinear (counter-propagating) laser beam. While this enables studying isotopes with lifetimes of a few ms, this also implies that the measured resonance frequency $\omega_\mathrm{col/acol}$ is Doppler-shifted from the rest-frame transition frequency $\omega_0$
\begin{eqnarray}
     \omega_\mathrm{col/acol} = \omega_0 \cdot \gamma (1\pm\beta)\mathrm{,}
\end{eqnarray}
where $\gamma=1/\sqrt{1-\beta^2}$ is the Lorentz factor, $\beta = v/\mathrm{c}$, $v$ is the speed of the ions in the laboratory frame, and $\mathrm{c}$ is the speed of light. This velocity dependence of the laboratory-frame resonance condition leads to a broadening of the measured line shape (Doppler broadening), obtained by folding the intrinsic line shape of the transition with the ion velocity distribution, mainly given by the thermal velocity distribution in the ion source. 
In CLS, Doppler broadening is strongly reduced by accelerating the ions of mass $m$ and charge $q$ through an electrostatic acceleration voltage $U_\mathrm{acc}$ of typically 10–60 keV. This compresses the Doppler width $\delta\omega_\mathrm{D}\simeq \frac{\omega_0}{\mathrm{c}}\frac{\delta E}{\sqrt{2mqU_\mathrm{acc}}}$ by several orders of magnitude, enabling measurements close to the limit imposed by the natural linewidth of allowed dipole transitions, typically in the $10–100\,$MHz range \cite{anton1978collinear}. However, this so-called Doppler compression does not suffice to reduce the Doppler width to the linewidth of a Raman transition, which is typically in the $100\,$kHz regime.

\subsection{\label{ssec:RamanTrans}Raman Transitions}

\begin{figure}
    \centering
    \includegraphics[width=0.9\linewidth]{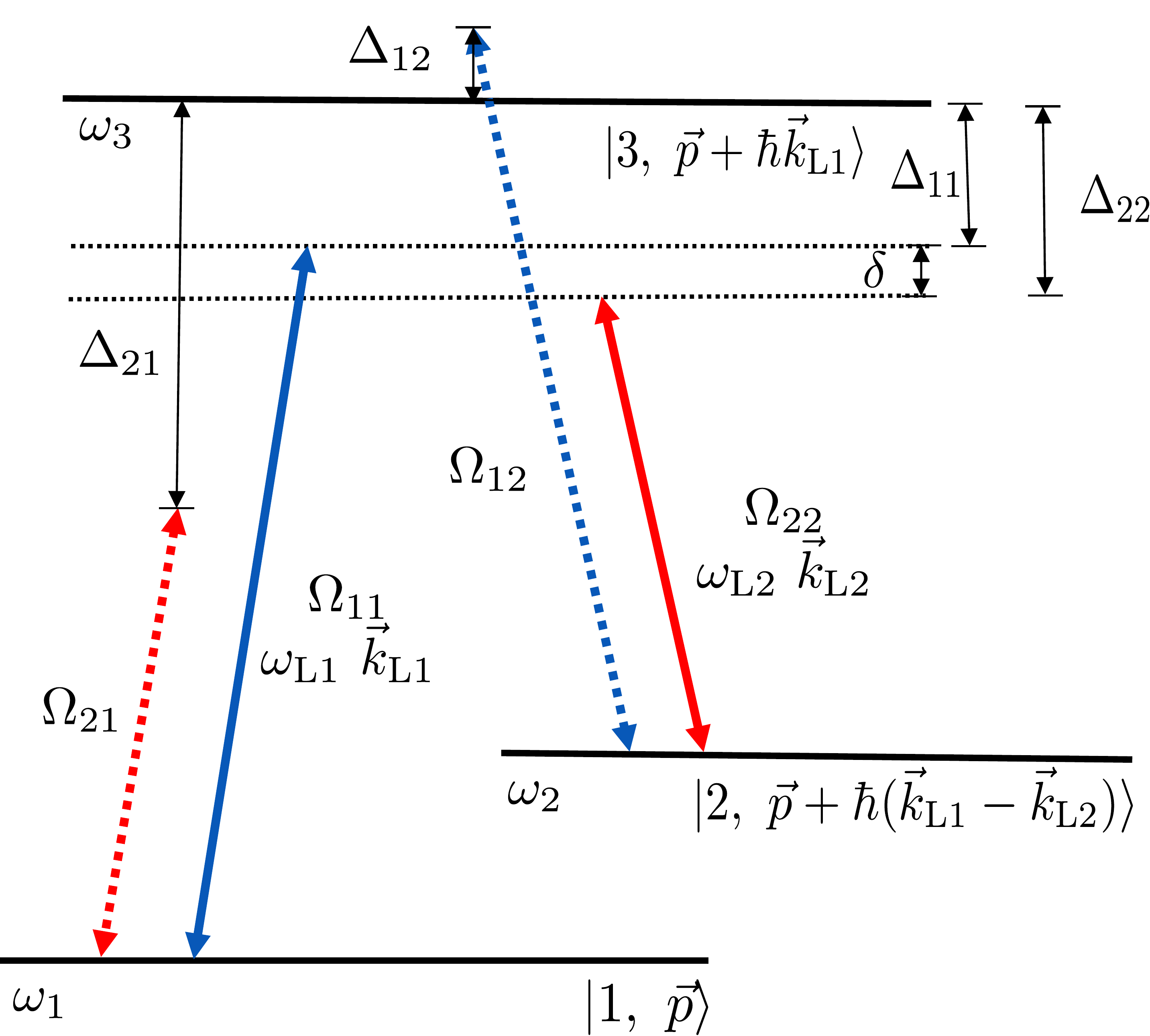}
    \caption{\label{fig:RamanLvlscheme}
    Raman transition in a three-level $\Lambda$-scheme. The arrows indicate the coupling of the laser $j \in \{1,2\}$ (frequency $\omega_{\mathrm{L}j}$ , wavevector $\Vec{k}_{Lj}$) to the dipole transition between the levels $\ket{n}$, $n \in \{1,2\}$ and $\ket{3}$ with the Rabi frequency $\Omega_{nj}$ and the single-photon detuning $\Delta_{nj}$. The full blue and red arrows indicate the couplings driving the Raman transition, while the dashed arrows correspond to the interactions that only result in additional AC-Stark shifts. Also indicated are the ion momenta in the different basis states, which have been included to account for Doppler-shifts and photon recoils.}
    \label{fig:enter-label}
\end{figure}
 
Figure \ref{fig:RamanLvlscheme} illustrates a Raman transition in a three-level $\Lambda$-scheme interacting with two lasers:
A first laser (frequency $\omega_{\mathrm{L}1}$, wave vector $\Vec{k}_{\mathrm{L}1}$) couples to a first dipole transition from an initial state $\ket{1}$ ($\hbar\omega_1$), here the ground state, to the intermediate state $\ket{3}$ ($\hbar\omega_3$).
The second laser ($\omega_{\mathrm{L}2}$, $\Vec{k}_{\mathrm{L}2}$) couples to a second dipole transition between a metastable state $\ket{2}$ ($\hbar\omega_2$) and the intermediate state $\ket{3}$. In such a system, a stimulated Raman transition is obtained by detuning both lasers from their dipole resonances by an amount $\Delta_{11}$ and $\Delta_{22}$, respectively. If the detuning of both lasers is identical, ergo the two-photon detuing $\delta=\Delta_{11}-\Delta_{22}$ is zero, this results in a direct two-photon transition from the ground state to the metastable state. 
This fits the intuitive picture of a transition from $\ket{1}$ to $\ket{2}$ through a virtual intermediate state, which is detuned from the real intermediate state $\ket{3}$ by $\Delta = \Delta_{11} = \Delta_{22}$. In Fig.\,\ref{fig:RamanLvlscheme} the virtual intermediate state is indicated by a horizontal dotted line and the coupling of $\ket{1}$ and $\ket{2}$ to this state via the detuned lasers by the blue and red full arrows, respectively. 
In resonance, photon of frequency $\omega_{\mathrm{L}1}$ is absorbed and simultaneously a photon of frequency $\omega_{\mathrm{L}2}$ is emitted through stimulated emission.\\
The far-off resonant coupling of the first laser to the second dipole transition and vice versa are indicated by blue/red dashed arrows and result in additional AC-Stark shifts, see Appendix \ref{secAppendA}. 
Neglecting spontaneous decay and assuming a classical electric field for the lasers ($\Vec{E_j} =  \frac{1}{2} \Vec{E}_{0,j}\, e^{i(\Vec{k}_{\mathrm{L}j}\cdot\Vec{r}-\omega_{\mathrm{L}j}t+\phi_j)}+ c.c.$), the Hamiltonian of such a system, using minimal coupling and applying the Power-Zienau-Woolley transformation as well as the rotating wave approximation in the interaction frame is given by \cite{dunning2015coherent}

\begin{equation}
    \begin{split}
        \hat{H}_\mathrm{IF} 
        &= \begin{pmatrix}
            0 & 0 & \frac{\hbar\Omega_{11}}{2} e^{-i\Delta t}\\
            0 & 0 & \frac{\hbar\Omega_{22}}{2} e^{-i(\Delta+\delta)t}\\
            \frac{\hbar\Omega_{11}^*}{2} e^{i\Delta t} & \frac{\hbar\Omega_{22}^*}{2} e^{i(\Delta+\delta)t}\mkern-54mu & 0\\
        \end{pmatrix}.
    \end{split}
    \label{eq:H_IF}
\end{equation}

The basis states are sorted such that $(a\quad b\quad c)^T=a\ket{1}+b\ket{2}+c\ket{3}$.
$\Omega_{nj} = -\bra{n}\hat{{d}} \cdot \Vec{E}_j \ket{3}/\hbar$ is the Rabi frequency of the dipole coupling of the level $\ket{n}$ to the level $\ket{3}$ through the laser $j$, with the dipole operator $\hat{{d}}$ and $\Delta_{nj}$ the single-photon detuning of that interaction. Far off-resonant couplings ($n\neq j$) have been omitted, as they only result in an additional AC-Stark shift. 
$\Delta$ is the detuning of the virtual intermediate state, given by
\begin{equation}
    \Delta := \Delta_{11} = \omega_{13}-(\omega_{\mathrm{L}1}-\frac{\Vec{p}\cdot\Vec{k}_{\mathrm{L}1}}{M}-\frac{\hbar k_{\mathrm{L}1}^2}{2M})\mathrm{,}\\
    \label{eq:Delta}
\end{equation}
with the dipole transition frequency $\omega_{13}=\omega_{3}-\omega_{1}$ and the ion momentum $\Vec{p}$. The term linear in $\Vec{k}_{\mathrm{L}1}$ is the non-relativistic Doppler shift, and the quadratic term is the energy shift induced by the absorption of the photon momentum.
$\delta$ is the two-photon detuning, given by
\begin{eqnarray}
    \delta = \Delta_{22} - \Delta_{11} =&& (\omega_{23}-\omega_{13})-(\omega_{\mathrm{L}2}-\omega_{\mathrm{L}1}\nonumber\\
    -\frac{\Vec{p}\cdot(\Vec{k}_{\mathrm{L}2}-\Vec{k}_{\mathrm{L}1})}{M} 
    &&+ \frac{\hbar}{2M}(\Vec{k}_{\mathrm{L}1}-\Vec{k}_{\mathrm{L}2})^2)\mathrm{.}
    \label{eq:delta}
\end{eqnarray}

Reducing the three-level system to an effective two-level system, see Appendix \ref{secAppendA}, reveals that for large $\Delta$ the population of the metastable state is given by
\begin{eqnarray}
    \label{eq:P2(t)}
    P_{2}(t) =&& \frac{\Omega_{\mathrm{R}}^2}{\Omega_\mathrm{R}^2 + (\delta-\delta_\mathrm{AC})^2}\nonumber\\
    &&\cdot\sin^2{\left(\frac{1}{2}\sqrt{\Omega_{\mathrm{R}}^2 + (\delta-\delta_\mathrm{AC})^2} t\right)}
\end{eqnarray}
if a pure ground-state population at $t=0$ is assumed.
This corresponds to a Rabi oscillation between the ground state $\ket{1}$ and the metastable state $\ket{2}$ with the two-photon Rabi frequency $\Omega_{\mathrm{R}} = \frac{\Omega_{11}\Omega_{22}^*}{2\Delta}$. The resonance condition is now given by the two-photon resonance condition $\delta-\delta_\mathrm{AC} = 0$, where $\delta_\mathrm{AC} = \Omega_1^\mathrm{AC}-\Omega_2^\mathrm{AC}$ is the difference in AC-Stark shifts of the first and second dipole transition induced by both lasers. For a large detuning and including far-off-resonant couplings, the AC-Stark shift is given in first order by
\begin{equation}
    \Omega_n^{AC} = \sum_{j=1,2}\frac{|\Omega_{nj}|^2}{4\Delta_{nj}}, \quad n \in \{1,2\}, 
\end{equation}
where $\Delta_{nj}$ is the detuning of the laser $j$ from the dipole transition from the state $\ket{n}$ to the metastable state $\ket{3}$.\\

\section{\label{sec:ExpSet}Experimental Setup}
\begin{figure}
    \centering
    \includegraphics[width=.8\linewidth]{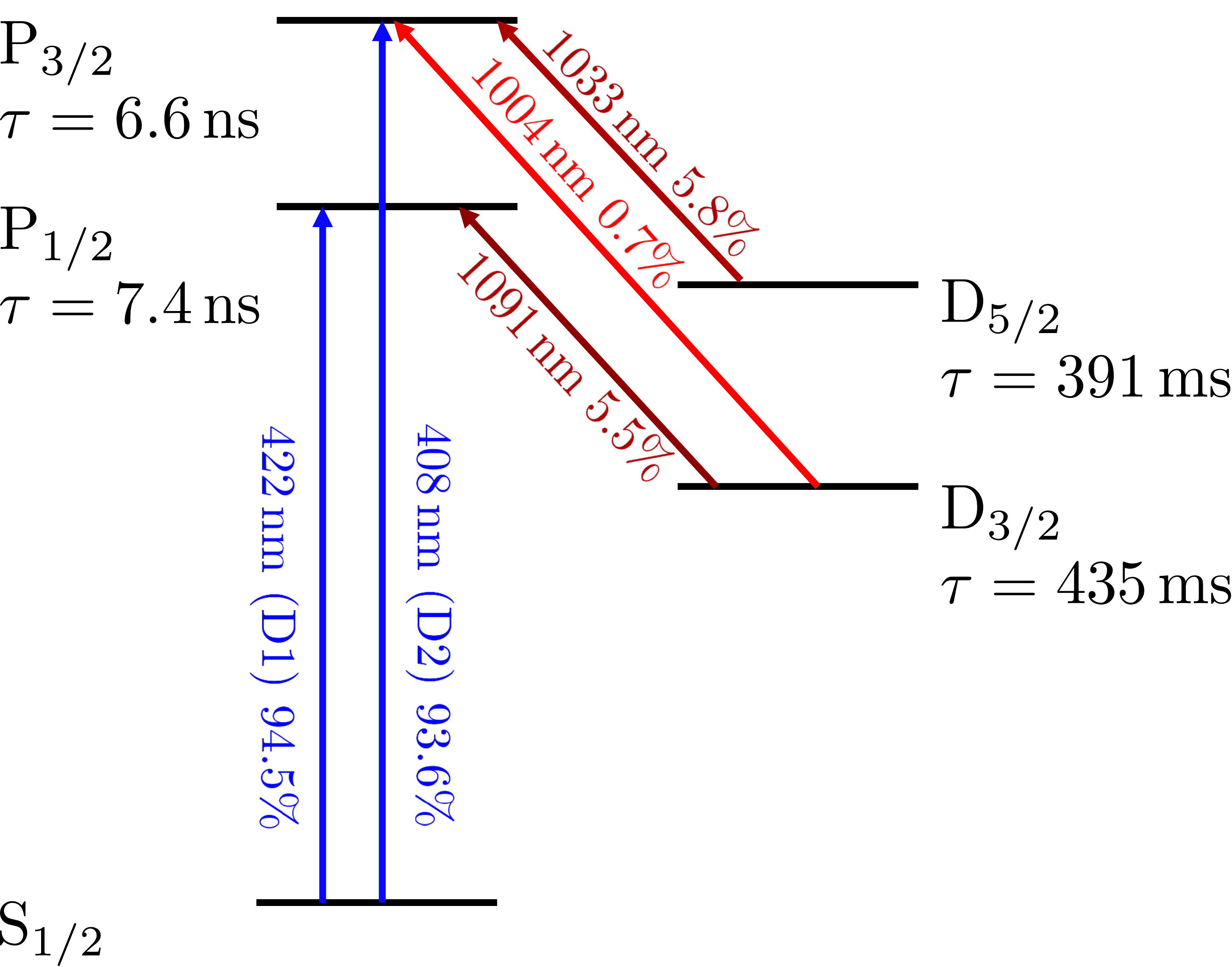}
    \caption{Fine structure level scheme of Sr$^+$. The arrows indicate different dipole transitions with their respective wavelength. Also indicated are life times $\tau$ of the different states and the branching ratios of the different decay paths. The life times and branching ratios were taken from \cite{NIST_ASD}.}
    \label{fig:Sr+LvlScheme}
\end{figure}

\begin{figure*}
    \centering
    \includegraphics[width=1\linewidth]{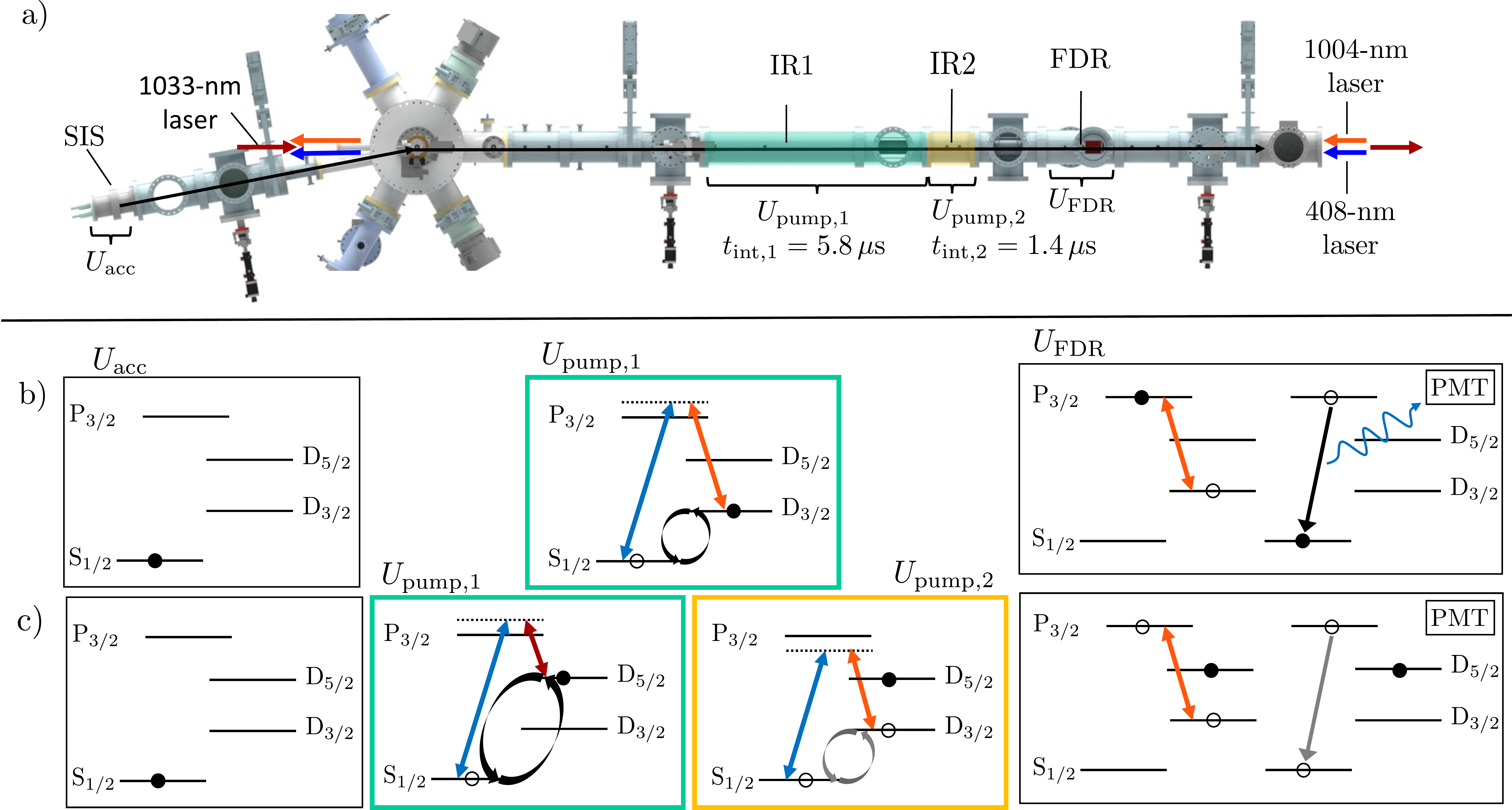}
    \caption{
    a) Schematics of the COALA beamline. The ions are produced in the surface ionization source (SIS) placed on a high voltage $U_\mathrm{acc}=20\,$kV, accelerated to ground potential and superimposed with the laser beams using electrostatic ion optics. All lasers are operated at a fixed frequency. Two interaction regions, the dedicated drift tube (IR1) and an einzel lens (IR2), are used to drive Raman transitions via Doppler tuning by floating the interaction regions to $U_\mathrm{pump,1}$ and $U_\mathrm{pump,2}$ respectively. The fluorescence detection region (FDR) is equipped with photomultiplier tubes, used to detect the 408-nm photons from the $\mathrm{P}_{3/2}\rightarrow\mathrm{S}_{1/2}$ decay.
    b) Measurement scheme for a single Raman transition, shown for the $\mathrm{S}_{1/2}\rightarrow\mathrm{D}_{3/2}$ transition. Initially, the ground state is populated. Spectroscopy is performed on the Raman transition via Doppler tuning by scanning $U_\mathrm{pump,1}$. A fixed voltage $U_\mathrm{FDR}$ is applied to the FDR to probe the $\mathrm{D}_{3/2}$ state by resonantly driving $\mathrm{D}_{3/2}\rightarrow\mathrm{P}_{3/2}$ dipole transition and detecting the 408-nm photons emitted from the subsequent $\mathrm{P}_{3/2}\rightarrow\mathrm{S}_{1/2}$ decay. 
    c) Measurements scheme for Doppler-free collinear Raman spectroscopy, shown for a $\mathrm{S}_{1/2}\rightarrow\mathrm{D}_{5/2}$ and a consecutive $\mathrm{S}_{1/2}\rightarrow\mathrm{D}_{3/2}$ transition. In a first Raman transition, driven at fixed $U_\mathrm{pump,1}$, ions of one velocity are selected and transferred to the $\mathrm{D}_{5/2}$ state. To perform a Doppler-free measurement the interaction potential $U_\mathrm{pump,2}$ of a second $\mathrm{S}_{1/2}\rightarrow\mathrm{D}_{3/2}$ Raman transition is scanned. The second transition is detected by probing the $\mathrm{D}_{3/2}$ state like in b). Ions of the velocity transferred out of the ground state in the first interaction are not transferred in the second interaction, as indicated by the grayed out transitions. Thus they do not contribute to the PMT signal, resulting in a Doppler-free Lamb dip in the resonance spectrum. 
    }
    \label{fig:Beamline+MSchemes}
\end{figure*}

The measurements were performed at the \underline{Co}llinear \underline{A}pparatus for \underline{L}aser Spectroscopy and \underline{A}pplied Science (COALA) at the Institute for Nuclear Physics at TU Darmstadt \cite{konig2020new}. A surface ionization source (SIS) was used to produce a Sr$^+$ beam of a few nA by heating a graphite crucible filled with Sr to about $2500\,$K via Ohmic heating. A level-scheme of Sr$^+$, including the relevant fine structure states, is shown in Fig.\,\ref{fig:Sr+LvlScheme}.
\\
Two Ti:Sa lasers (Sirah Matisse-2), pumped by frequency-doubled, diode-pumped $\mathrm{Nd:YVO}_4$ lasers (Spectra-Physics Millennia eV), are used to produce the $408$-nm and $1004$-nm continuous-wave laser beams. The $1004$-nm light is directly produced in the Ti:Sa laser. A frequency doubling stage (Spectra-Physics Wavetrain) equipped with an LBO crystal is used in combination with the second Ti:Sa laser at $816\,$nm to produce $408$-nm light. The latter is guided through an acousto-optic modulator (AOM), as required for the implemented measurement scheme, see Sec.\,\ref{sec:MScheme}. 
These two laser beams are superimposed using a long-pass dichroic mirror and guided through the beamline in anticollinear geometry.
Each Ti:Sa laser is short-term stabilized to a reference cavity, which is long-term stabilized to a frequency comb (Menlo Systems FC1500-250-WG). In combination with a wavemeter (HighFinesse WS8-2), this allows frequency measurements at sub $50$-kHz accuracy, limited by the laser linewidth.
In addition, a tuneable diode laser at $1033\,$nm (Toptica DL Pro) is used incollinear geometry. Its frequency is stabilized to a second wavemeter (High Finesse WSU 30) via a PID loop and monitored in parallel with the WS8-2. Both wavemeters are stabilized to the same He:Ne-laser. As measuring all three laser frequencies simultaneously using the available frequency comb is not possible, measurements of the diode laser frequency using frequency-comb are performed regularly, and used to reference the wavemeter reading. The employed wavemeters are known to have a frequency-dependent offset \cite{konig2020WavemeterPerformance}. The frequency comb measurements are used to determine that offset. which is linearly interpolated between reference measurements to correct for drifts due to, \textit{e.g.}, temperature changes.
\\
The ion source is placed on a $U_\mathrm{acc}=20\,$kV potential, and the ions are accelerated toward the ground potential of the beamline shown in Fig.\,\ref{fig:Beamline+MSchemes} a). Electrostatic steering electrodes and a quadrupole doublet are used to bend the ion beam by $10\,^{\circ}$ into the main beamline, optimize the ion beam profile, and align the ion beam with the laser beams. The laser/ion overlap is ensured using multi-channel plates (MCPs) followed by  phosphorus screens. The MCP-screen combination can be lowered into the beam, and the beam spot is observed with a camera. A 1.2-m-long section in the center of the beamline is equipped with a tube that can be floated to a separate voltage $U_\mathrm{pump, 1}$, allowing for a well-defined interaction region and interaction potential for the Raman transition. An einzel lens behind this section can be used as a second 0.3-m-long interaction region by floating all three of its electrodes to a second interaction voltage $U_\mathrm{pump, 2}$ for an optional subsequent second Raman transition.\\
The fluorescence detection region (FDR) located at the end of the beamline is equipped with photomultiplier tubes (PMTs) to detect fluorescence light in the UV range. The PMT counts are fed into the time-resolved TILDA data acquisition system \cite{phdSKaufmann}, which is based on two field-programmable gate arrays (FPGA). This system is also used to scan the laser frequency or the voltage
applied to the interaction region and to trigger the AOM of the 408-nm-laser. \\

\section{\label{sec:MScheme}Measurement Scheme}

As an example, the measurement scheme for the $\mathrm{S}_{1/2}\rightarrow\mathrm{D}_{3/2}$ Raman transition is depicted in Fig.\,\ref{fig:Beamline+MSchemes} b). The scheme for the  $\mathrm{S}_{1/2}\rightarrow\mathrm{D}_{5/2}$ transition works analogously.
Instead of scanning the laboratory-frame laser frequencies to match the two-photon resonance condition of the Raman transition, the voltage of the interaction region $U_\mathrm{pump,1}$ is scanned. This modifies the ion velocity and thus, the Doppler-shifted laser frequencies in the ion rest frame. Since the Doppler shift depends on the frequency and the direction of the laser, the laser frequencies shift differently, allowing to scan the two-photon detuning of the Raman transition.\\
As driving the $\mathrm{S}_{1/2}\rightarrow\mathrm{D}_{J}$ ($\mathrm{J}\in\{3/2,5/2\}$) Raman transition leads to an increase in the population of the $\mathrm{D}_{J}$ state, detection of the Raman transition is performed by continuously probing the $\mathrm{D}_{J}$ state via the $\mathrm{D}_{J}\rightarrow\mathrm{P}_{3/2}$ dipole transition. This is achieved by applying a fixed voltage $U_\mathrm{FDR}$ to the FDR, shifting the Doppler-shifted laser frequency in the ion rest frame into resonance. The ions that were transferred into the $\mathrm{D}_{J}$ state in the interaction region are thus excited into the $\mathrm{P}_{3/2}$ state in the FDR. They subsequently decay into the $\mathrm{S}_{1/2}$ ground state, emitting 408-nm-photons, resulting in an increase in PMT counts. During the probing in the FDR, the 408-nm laser is turned off using the AOM to reduce the laser-induced background.
\\
This scheme is susceptible to Doppler-broadening. Hence, for typical ion energy widths of our ion source of a few eV, the line shape of the detected Raman resonance is predominantly given by the energy distribution of the ground-state population. To develop a narrow-width velocity filter for ion beams using Raman transitions, a scheme employing two sequential Raman transitions was tested: First, an  $\mathrm{S}_{1/2}\rightarrow\mathrm{D}_{5/2}$ Raman transition is driven in the first interaction region at a fixed interaction potential $U_\mathrm{pump,1}$. Analogously to saturation spectroscopy, this results in a Doppler-free Lamb dip in the ground-state energy distribution, corresponding to the velocities selected/filtered by the first Raman transition. This dip is then probed by either a second, consecutive $\mathrm{S}_{1/2}\rightarrow\mathrm{D}_{3/2}$ Raman transition, as shown in Fig.\,\ref{fig:Beamline+MSchemes} c), or  $\mathrm{S}_{1/2}\rightarrow\mathrm{D}_{5/2}$ Raman transition. This is realized by scanning the voltage $U_\mathrm{pump,2}$ applied to the second interaction region. 
Alternatively, the peak in the velocity distribution of the $\mathrm{D}_{5/2}$ population can be probed via a  $\mathrm{D}_{5/2}\rightarrow\mathrm{D}_{3/2}$ Raman transition, which opens up the possibility of a first direct practically Doppler-free measurement of the $\mathrm{D}_{5/2}\rightarrow\mathrm{D}_{3/2}$ transition frequency. 
The excitation to the $\mathrm{D}_\mathrm{J}$ is tested by resonantly probing the metastable state via the  $\mathrm{D}_\mathrm{J}\rightarrow\mathrm{P}_{3/2}$ dipole transition and detecting the photons from the $\mathrm{P}_{3/2}\rightarrow\mathrm{S}_{1/2}$ decay, like for a single Raman transition.
Applying different interaction potentials $U_\mathrm{pump,1}$, $U_\mathrm{pump,2}$, and $U_\mathrm{FDR}$ in particular allows the separation of the different Raman transitions without having to use additional lasers or AOMs.
\\    

\section{\label{sec:Simulations}Simulations}
\subsection{Method}
To test the feasibility of performing collinear Raman spectroscopy and using Raman transitions as a velocity filter at COALA, simulations similar to the ones performed by Neumann \textit{et al.}~\cite{neumann2020raman} were performed specifically for an $^{88}$Sr$^+$  beam at $20\,$keV.\\
The state populations in the $^{88}$Sr$^+$ $\mathrm{S}_{1/2}$, $\mathrm{P}_{3/2}$, $\mathrm{D}_{3/2}$, $\mathrm{D}_{5/2}$ $\Lambda$-scheme were calculated 
using the interaction-frame Hamiltonian of this four-level system, 

\begin{widetext}
    \begin{equation}
    \begin{split}
        \hat{H}_\mathrm{IF}
        &= \begin{pmatrix}
            0 & 0 & 0 &\frac{\hbar\Omega_{11}}{2} e^{-i(\Delta t+\phi_1)}\\
            0 & 0 & 0 &\frac{\hbar\Omega_{22}}{2} e^{-i((\Delta+\delta_{12})t+\phi_2)}\\
            0 & 0 & 0 &\frac{\hbar\Omega_{33}}{2} e^{-i((\Delta+\delta_{13})t+\phi_3)}\\
            \frac{\hbar\Omega_{11}^*}{2} e^{i(\Delta t+\phi_1)} & \frac{\hbar\Omega_{22}^*}{2} e^{i((\Delta+\delta_{12})t+\phi_2)} & \frac{\hbar\Omega_{33}^*}{2} e^{i((\Delta+\delta_{13})t+\phi_2)} & 0\\
        \end{pmatrix}\mathrm{.}
    \end{split}
    \label{eq:Hii_doubleR}
\end{equation}
\end{widetext}

The same notation as in Eq.\,\eqref{eq:H_IF} is used and a phase $\phi_j$ of the laser $j$ is included. The indices $n$ in the basis states as well as in the single-photon Rabi frequencies $\Omega_{nj}$ and detunings $\Delta_{nj}$ are such that $\{1,2,3,4\}\hat{=}\{\mathrm{S}_{1/2},\,\mathrm{P}_{3/2},\mathrm{D}_{3/2},\,\mathrm{D}_{5/2}\}$. Laser 1 is the anticollinear 408-nm laser, laser 2 the anticollinear 1004-nm laser, and laser 3 the collinear 1033-nm laser.
$\Delta$ is given by Eq.\,\eqref{eq:Delta}, except for the index shift for the metastable state, and $\delta_{1n}=\Delta_{nn} - \Delta_{11}$ $n \in \{2,3\}$ is defined according to Eq.\,\eqref{eq:delta}. Again, far off-resonant couplings and spontaneous decay were omitted. The Doppler shifts in Eqs.\,\eqref{eq:Delta} and \eqref{eq:delta} were replaced by the corresponding relativistic terms. Eliminating the time dependencies of this Hamiltonian by performing a unitary transformation with 

\begin{equation}
    \hat{U}= \begin{pmatrix}
            1 & 0 & 0 & 0\\
            0 & e^{i\delta_{12}t} & 0 & 0\\
            0 & 0 & e^{i\delta_{13}t} & 0\\
            0 & 0 & 0 & e^{-i\Delta t}\\
        \end{pmatrix}
\end{equation}
yields
\begin{equation}
    \hat{H}_{4}= \begin{pmatrix}
            0 & 0 & 0 & \frac{\hbar\Omega_{11}}{2} e^{-i\phi_{L1}}\\
            0 & -\hbar\delta_{12} & 0 & \frac{\hbar\Omega_{22}}{2} e^{-i\phi_{L2}}\\
            0 & 0 & -\hbar\delta_{13} & \frac{\hbar\Omega_{33}}{2} e^{-i\phi_{L3}}\\
            \frac{\hbar\Omega_{11}^*}{2} e^{i\phi_{L1}} & \frac{\hbar\Omega_{22}^*}{2} e^{i\phi_{L2}} & \frac{\hbar\Omega_{33}^*}{2} e^{i\phi_{L3}} & \hbar\Delta\\
        \end{pmatrix}\mathrm{.}
    \label{eq.:H_tind}
\end{equation}
The time evolution of an initial state $\ket{\Phi}_0=\ket{\Phi}(t=0)$ is then calculated using the ansatz
\begin{equation}
    \label{eq:TimeEvol}
     \ket{\Phi}(t) = e^{-i\hbar\hat{H}_{4}t}\mathrm{.}
\end{equation}

Assuming Gaussian laser beam profiles with a beam waist $w_j$ and a total power $P_j$, the Rabi frequencies are obtained using
\begin{equation}
    |\Omega_{nj}| = \frac{|d_{\mathrm{eff}, n}|}{\hbar}\sqrt{\frac{4P_j}{\pi\epsilon_0 \mathrm{c} w_j^2}},
\end{equation}
where $\epsilon_0$ is the vacuum permittivity.
$|d_{\mathrm{eff}, n}|$ is the effective dipole moment of the transition $n$. Neglecting hyperfine splitting, as $I=0$ for $^{88}\mathrm{Sr}$, and magnetic substates, and hence polarization, the effective dipole moment can be calculated with the reduced dipole matrix element $\langle J \| e\hat{\vec{r}} \| J' \rangle$ 
\begin{equation}
    \label{eq:el_dip}
    |d_{\mathrm{eff}, n}|^2 = \frac{1}{3}|\langle J \| e\hat{\vec{r}} \| J' \rangle|^2\mathrm{,}
\end{equation}
where $J$ and $J'$ are the angular momentum quantum numbers of the initial and final state of the transition $n$. This reduced matrix element can be extracted from the relation 
\begin{equation}
    A_{J'J}=\frac{1}{\tau} = \frac{\omega_0^3}{3\pi \epsilon_0 \hbar \mathrm{c}^3}\frac{2J+1}{2J'+1} |\langle J \| e\hat{\vec{r}} \| J' \rangle|^2,
\end{equation}
using the Einstein coefficient $A_{J'J}$ of the transition or its lifetime $\tau$ as well as its transition frequency $\omega_0 = 2\pi\nu_0$. 
The literature values used to calculate the Rabi frequencies in the simulations discussed below are listed in Tab.\,\ref{tab:params}.
\\

Both the ion energy distribution and the spatial intensities of all three laser beams and the ion beam are included in the simulations. The ion energy is presumed to follow a Gaussian distribution with a standard deviation $\sigma_E$ and is included by folding the beam-energy-dependent state population with the energy distribution. 
The spatial intensity distributions are taken into account by generating a two-dimensional grid in coordinate space representing the beam cross-section. Gaussian intensity distributions of beam waists $w_i$ are inferred for laser and ion beams based on measurements of the different beam profiles. All beams are assumed to be perfectly aligned and collimated.
The laser beam intensities are fixed for each grid point, and the time evolution of the system is then calculated.
For sequential interactions, the final state and energy distributions after the first interaction are taken as input for the simulation of the second interaction.
The maximum expected interaction times $ t_\mathrm{interact, 1/2}$ were calculated from the beam energy and the lengths of the interaction regions and are $5.8\,\mu$s and $1.4\,\mu$s for the first and second interaction regions, respectively. For each grid point, the state populations are weighed with the ion-beam intensity distribution.
The total population of the beam is obtained by integrating over the beam cross-section and the energy distribution.
\\
The total laser beam powers were chosen to be realistic for the given experimental conditions and are listed with the different beam waists $w_i$ in Tab.\,\ref{tab:params}. The detuning of the virtual intermediate state was calculated based on the targeted phase of the two-photon Rabi oscillation in the beam center for the mean ion energy, and the laboratory frame laser frequencies were calculated from this detuning and the literature values for the dipole transition frequencies. For the simulations shown, the targeted phase is $3\pi$ for all interactions. Unless specified otherwise, the parameters in Tab.\,\ref{tab:params} were applied. The same strategy was used to find the initial parameters for the experimental campaign.\\

\begin{table*}
\caption{\label{tab:params}List of the experimental parameters and literature values for the parameters of the dipole transitions used for all simulations unless specified otherwise. The dipole transition frequencies are taken from \cite{palmes2025Sr}, the $A$-factors from \cite{NIST_ASD} and the mass $m$ from \cite{wang2021ame}}
\begin{ruledtabular}
\begin{tabular}{cccccc}
 parameter & value & parameter & value & parameter & value \\ \hline
 m (u) & 87.905063674 & q (e) & 1 &  & \\
 $\nu_{\mathrm{S}_{1/2}\rightarrow\mathrm{P}_{3/2}}$(MHz)&734 989 824&
 $\nu_{\mathrm{D}_{3/2}\rightarrow\mathrm{P}_{3/2}}$(MHz)&298 616 134&
 $\nu_{\mathrm{D}_{5/2}\rightarrow\mathrm{P}_{3/2}}$(MHz)&290 210 798\\
 $A_{\mathrm{S}_{1/2}\rightarrow\mathrm{P}_{3/2}}$($\mu\mathrm{s}^{-1}$)&141&
 $A_{\mathrm{D}_{3/2}\rightarrow\mathrm{P}_{3/2}}$($\mu\mathrm{s}^{-1}$)&1.0&
 $A_{\mathrm{D}_{5/2}\rightarrow\mathrm{P}_{3/2}}$($\mu\mathrm{s}^{-1}$)&8.7\\
 $U_\mathrm{acc}$(kV)&20&$U_\mathrm{pump,1}$(V)&250&$U_\mathrm{pump,2}$(V)&380\\
 $\nu_\mathrm{L1}$(MHz)&734 480 226.62&$\nu_\mathrm{L2}$(MHz)&298 408 493.56&$\nu_\mathrm{L3}$(MHz)&290 413 141.74\\
 $P_\mathrm{L1}$(mW)&1.5&$P_\mathrm{L2}$(mW)&400&$P_\mathrm{L3}$(mW)&10\\
 $w_\mathrm{L1}$(mm)&1.0&$w_\mathrm{L2}$(mm)&2.0&$w_\mathrm{L3}$(mm)&1.5\\
 $\sigma_E$ (meV)&500&$w_\mathrm{ion}$(mm)&5.0&&\\
 $\Omega_{\mathrm{R},\mathrm{S}_{1/2}\rightarrow\mathrm{D}_{5/2}}$(MHz)& 1.64 &$\Delta_{\mathrm{S}_{1/2}\rightarrow\mathrm{D}_{5/2}}$(MHz)& -2$\pi\cdot$708.89 &&\\
 $\Omega_{\mathrm{R},\mathrm{S}_{1/2}\rightarrow\mathrm{D}_{3/2}}$(MHz)& 2.19 &$\Delta_{\mathrm{S}_{1/2}\rightarrow\mathrm{D}_{3/2}}$(MHz)& 2$\pi\cdot$999.91&&
\end{tabular}
\end{ruledtabular}
\end{table*}

\subsection{\label{sim:velFilter}Velocity Filter}
\begin{figure}
    \centering
    \includegraphics[width=1.0\linewidth]{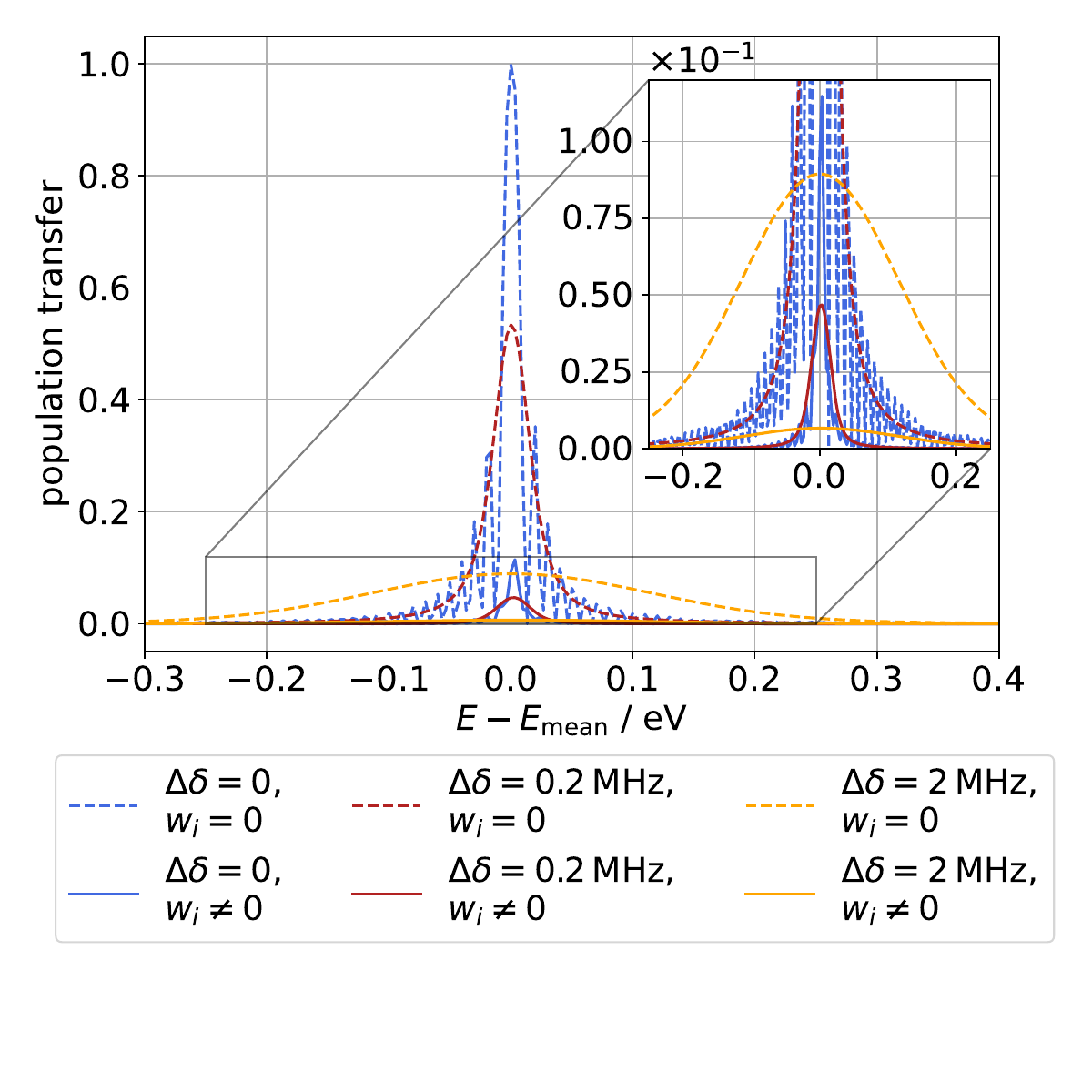}
    \caption{Simulations of the energy dependent population transfer for a $3\pi$ $\mathrm{S}_{1/2}\rightarrow\mathrm{D}_{5/2}$ Raman transition. The dotted lines indicate simulations without spatial intensity distribution ($w_i=0$), solid lines simulations with beam diameters according to Tab.\,\ref{tab:params} ($w_i\neq0$). The different colors show simulations for different variations in the two-photon detuning $\Delta \delta \in \{0,\, 0.2\,\mathrm{MHz},\, 2\,\mathrm{MHz}\}$, induced by according laser line-widths or variations in the interaction potential $\Delta U \in \{0,\, 0.01\,\mathrm{V},\,0.1\,\mathrm{V}\}$.}
    \label{fig:Sim_filter_eff}
\end{figure}

\begin{figure}
    \centering
    \includegraphics[width=.6\linewidth]{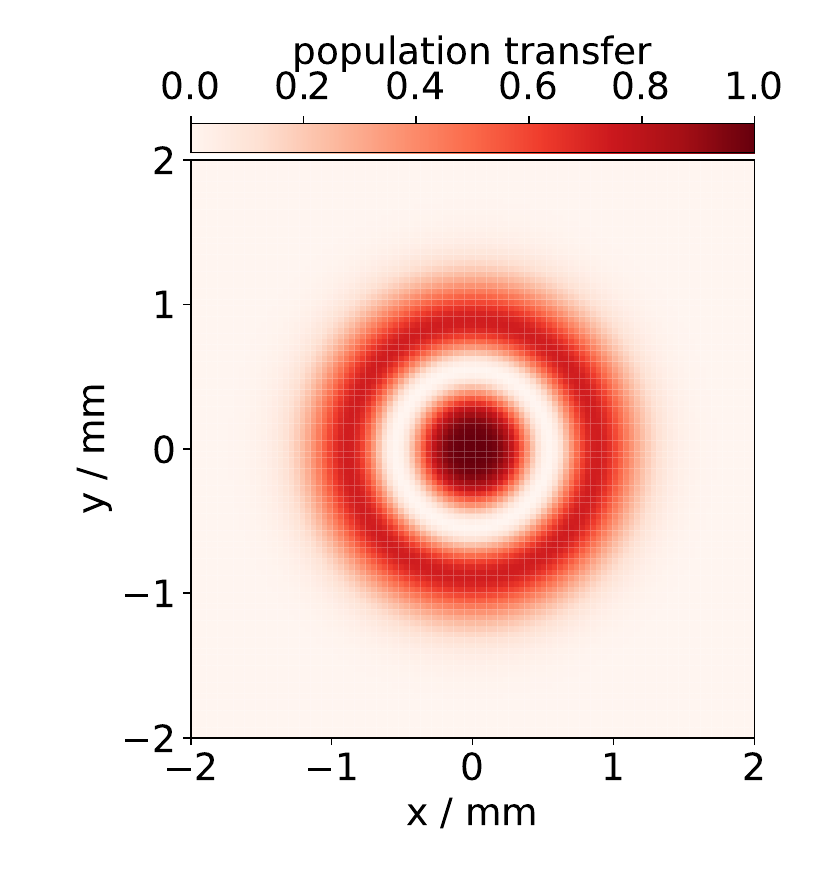}
    \caption{Simulated population transfer across the the ion-beam cross-section including spatial intensity distributions for a $3\pi$ $\mathrm{S}_{1/2}\rightarrow\mathrm{D}_{5/2}$ Raman transition in a Gaussian beam profile according to Tab.\,\ref{tab:params}. The phase of the Raman transition decreases with the distance from the beam center, reflecting the spatial intensity distribution of the laser beams, resulting in a local minimum ($2\pi$-pulse) and a second local maximum ($1\pi$-pulse). In the second maximum the population transfer does not reach $1$ due to a smaller AC-Stark shift compared to the beam center.}
    \label{fig:PopTrans2D}
\end{figure}

Figure \ref{fig:Sim_filter_eff} shows the simulated energy-dependent population transfer from the ground state into the metastable $\mathrm{D}_{5/2}$ state after a $3\pi$-pulse.\\
Comparing simulations which are including finite beam waists ($w_i\neq0$) with those that are excluding ($w_i=0$) spatial intensity distributions, shows a significant decrease in the efficiency of the Raman velocity filter in the former case. As the single-photon Rabi frequencies depend on the laser intensities, so does the phase of the Raman transition and therefore, the population transfer. This is visualized in Fig.\,\ref{fig:PopTrans2D}. Ions not located in the beam center do not experience the full $3\pi$-pulse, reducing the total population transfer. This effect is enhanced by the fact that for the beam diameters chosen in the experiment, the ion beam is significantly larger than the laser beams. Furthermore, a second local maximum in the population arises at a distance from the beam center at which the laser intensities match a $1\pi$-pulse. No broadening of the selected energy class due to the spatial variation of the laser intensities, and thus the AC-Stark shift, can be seen. However, the AC-Stark shift leads to a shift in the energy of the selected ions, since the AC-Stark shift averaged over the entire beam cross section ($w_i\neq0$) differs from the one at the beam center ($w_i=0$).\\
Both variations $\Delta U$ in the interaction potential and the finite laser line-widths lead to a variation $\Delta \delta$ in the ion rest frame. 
In Fig.\,\ref{fig:Sim_filter_eff} the energy-dependent population transfer is compared for $\Delta U \in \{0, 0.01\,\mathrm{V}, 0.1\,\mathrm{V}\}$, corresponding to $\Delta \delta \in \{0, 0.2\,\mathrm{MHz}, 2\,\mathrm{MHz}\}$. For finite $\Delta \delta$ reaching a total transfer efficiency of $1$ is no longer possible, even for negligible beam diameters. Moreover, the energy width of the ions selected by the Raman transition increases from $7\,$meV to $18\,$meV and $135\,$meV. Although for ion beams with energy widths $\sigma_E \gg \Delta U$ this does not affect the total population transfer, this is critical when implementing a Raman velocity filter, as discussed in Sec.\,\ref{ssec:VelFilter}.
\\

\subsection{Rabi Oscillations \& Sequential Raman Transitions}
\begin{figure}
    \centering
    \includegraphics[width=1.0\linewidth]{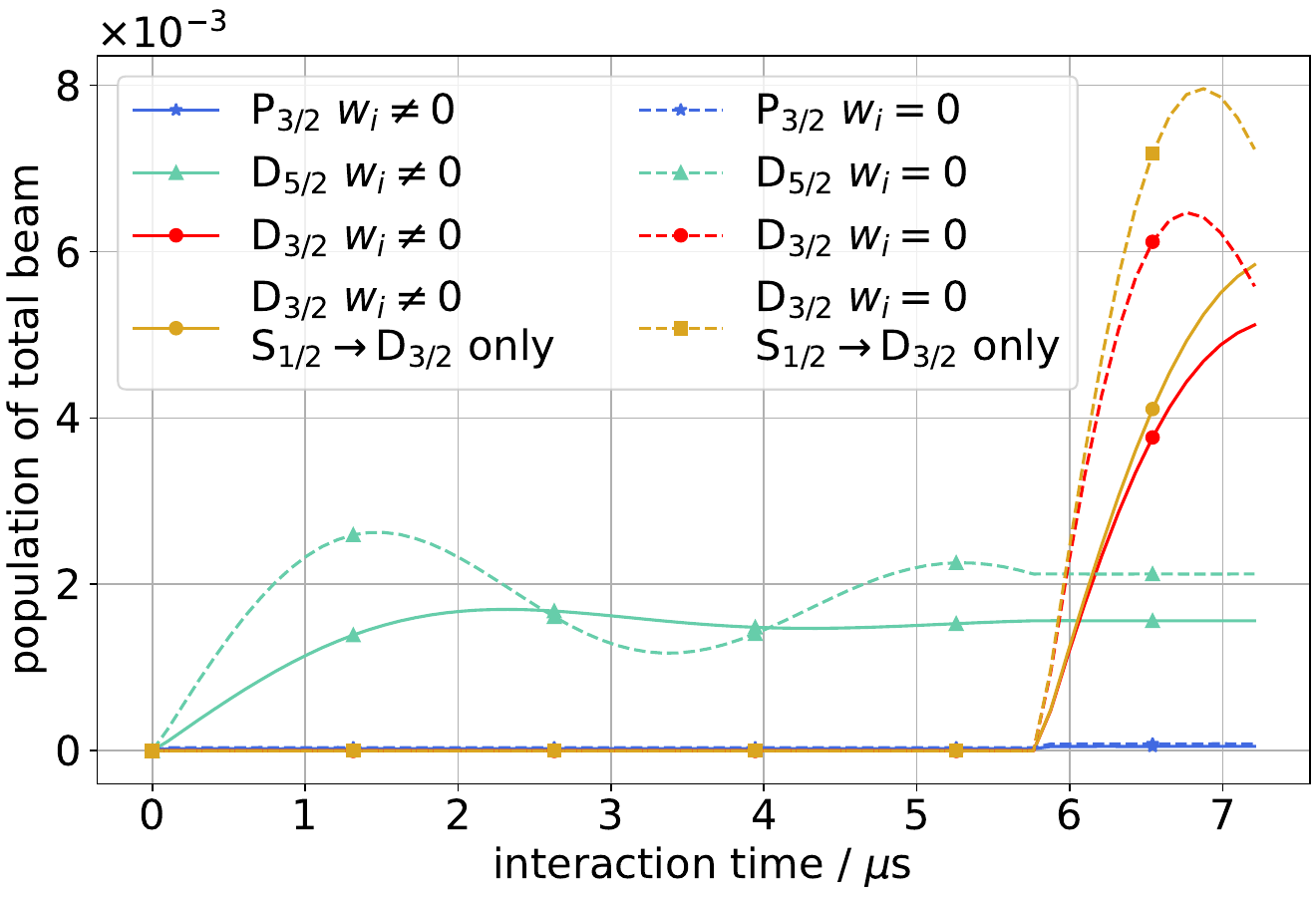}
    \caption{Time-dependent population transfer for a $3\pi$ $\mathrm{S}_{1/2}\rightarrow\mathrm{D}_{5/2}$ transition in the first interaction region ($0-5.8\,\mu$s) and a $1\pi$ $\mathrm{S}_{1/2}\rightarrow\mathrm{D}_{3/2}$ transition in the second interaction region ($5.8-7.2\,\mu$s) and only a $\mathrm{S}_{1/2}\rightarrow\mathrm{D}_{3/2}$ transition in the second interaction region. The solid lines indicate the population as fraction of the total beam including spatial intensity distributions of both laser and ion beam and the dashed lines the population when only including the ion energy distribution. The latter is scaled down by a factor of $10$ for better comparison.}
    \label{fig:Sim_Collapse_RO}
\end{figure}

Simulations on a $3\pi$ $\mathrm{S}_{1/2}\rightarrow\mathrm{D}_{5/2}$ and a consecutive $1\pi$ $\mathrm{S}_{1/2}\rightarrow\mathrm{D}_{3/2}$ Raman transition were performed to test the feasibility of applying the scheme for Doppler-free collinear Raman spectroscopy presented in Sec.\,\ref{sec:MScheme}. Figure \ref{fig:Sim_Collapse_RO} shows the time-dependent state populations obtained using the parameters in Tab.\,\ref{tab:params} for the sequential interactions, as well as for only the second interaction. The latter is included to investigate the strength of the Lamb-dip, which corresponds to the difference between the population transfer of only the $\mathrm{S}_{1/2}\rightarrow\mathrm{D}_{3/2}$ interaction and the population of the $\mathrm{D}_{3/2}$ state in the sequential scheme. The state populations were again calculated with ($w_i\neq0$) and without ($w_i=0$) spatial beam intensity distributions.
\\
As the two-photon Rabi frequency depends on the detuning $\Delta$ in the ion rest frame and thus the ion velocity and energy, 
the finite width of the ion energy distribution leads to a dampening of the two-photon Rabi oscillation. 
Since the total population transfer is given by the fold of the energy dependent Rabi oscillation and the ion-energy distribution, the effective oscillation of the population transfer is given by an average over oscillations of different frequencies. This averaging also results in a shift of the $3\pi$ maximum in the oscillation of the $\mathrm{D}_{3/2}$ state population from the $5.8\,\mu$s calculated for the mean ion energy to $5.3\,\mu$s.
\\
For an ion-energy width of $0.5\,$eV, this variation in two-photon Rabi frequencies is less than $2$\%, and thus, the oscillation is only dampened.
However, if additionally the spatial intensity distributions of the laser and ion beams are taken into account ($w_i\neq0$), this enhances this this effect. The averaging now includes the spatial dependence of the Rabi frequency. As the ion beam is larger than the laser beams, this yields contributions with frequencies ranging from the frequency in the beam center to almost zero. The variation of the phases of these contributions leads to a collapse of the Rabi oscillation in the accumulated signal.

\begin{figure}
    \centering
    \includegraphics[width=1\linewidth]{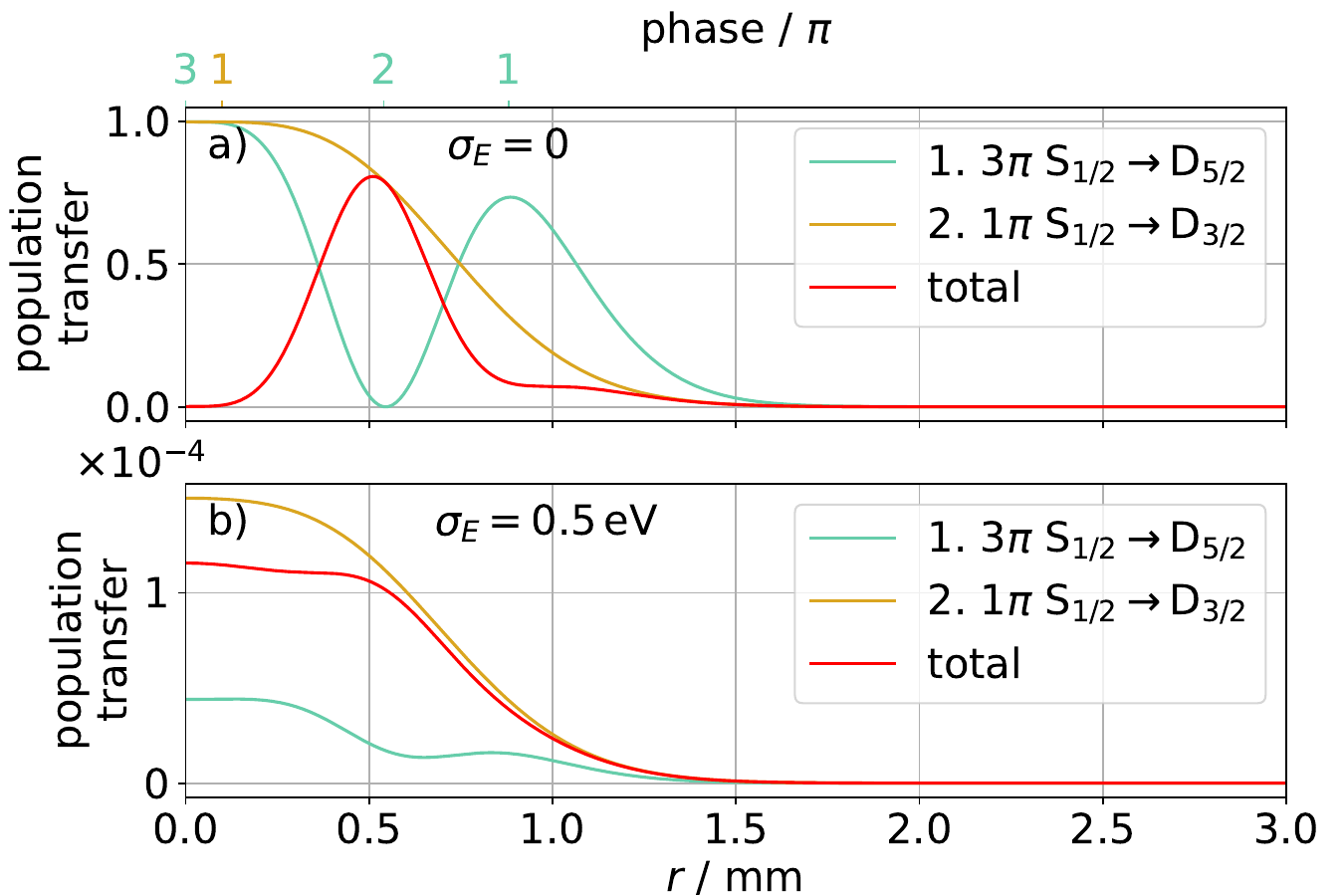}
    \caption{Radial dependence of the population transfer for a sequential $3\pi$ $\mathrm{S}_{1/2}\rightarrow\mathrm{D}_{5/2}$ (in interaction region 1 (IR1)) and a $1\pi$ $\mathrm{S}_{1/2}\rightarrow\mathrm{D}_{3/2}$ (IR2) Raman transition. Shown are the population transfer of the individual transitions and the total population transfer for ion energy beam widths of a) $\sigma_E=0$ and b) $\sigma_E=0.5\,$eV. In a) the upper $x$-axis indicates the phase of the individual transitions. 
    The difference between the $\mathrm{S}_{1/2}\rightarrow\mathrm{D}_{3/2}$ population transfer and the total population transfer corresponds to the strength of the Lamb dip. It is reduced if the population transfer of the first interaction decreases, \textit{e.g.} at a $2\pi$-pulse.  
    In b) the population transfer of the second transition is higher because its two-photon Rabi frequency is higher and the lasers are co-propagating, resulting in a broader selection in ion energy. This further reduces the strength of the Lamb dip}
    \label{fig:SimPopTransRadial}
\end{figure}

\begin{figure}
    \centering
    \includegraphics[width=1\linewidth]{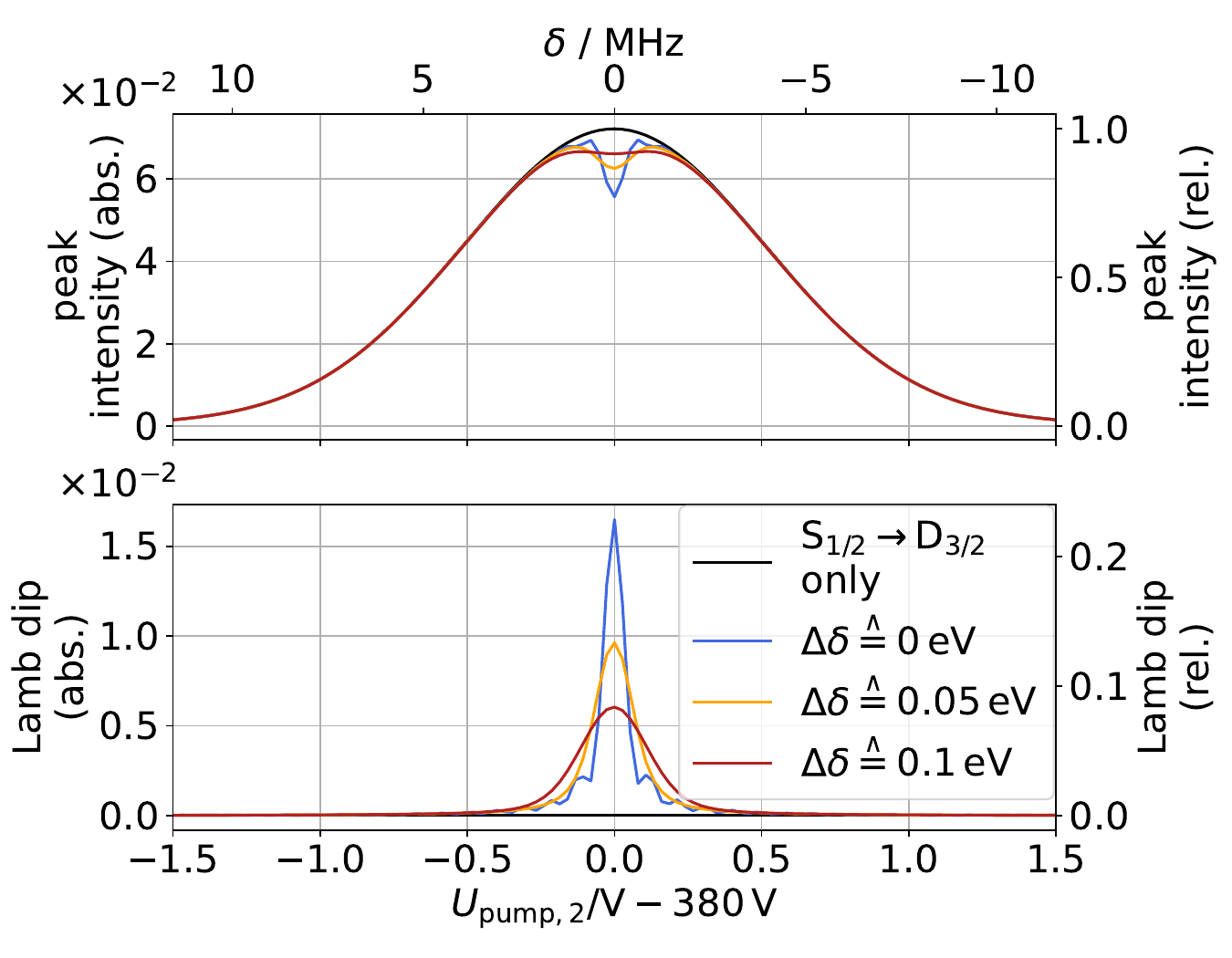}
    \caption{Simulated spectra obtained when scanning the interaction potential $U_\mathrm{pump,2}$ of a $1\pi$ $\mathrm{S}_{1/2}\rightarrow\mathrm{D}_{3/2}$ Raman transition after a $3\pi$ $\mathrm{S}_{1/2}\rightarrow\mathrm{D}_{5/2}$ Raman transition. 
    The upper half shows the resonance for different fluctuations of the interaction potential, see Fig.\,\ref{fig:Sim_filter_eff}, and the lower half shows the Lamb dip obtained from the difference in population transfer to just the $\mathrm{S}_{1/2}\rightarrow\mathrm{D}_{3/2}$ transition.
    The left and right $y$-axis shows the peak intensity as the fraction of the total ion beam in the $\mathrm{D}_{3/2}$ state and relative to the maximum population transfer of just the $\mathrm{S}_{1/2}\rightarrow\mathrm{D}_{5/2}$ transition, respectively.
    The upper $y$-axis indicates the two-photon detuning $\delta$.}
    \label{fig:Sim_Spectra}
\end{figure}

The strength of the Lamb dip is limited to $\sim20\%$. This is due to two factors:
Firstly, the different pulse lengths and laser beam diameters of the two interactions lead to different locations of the maxima and minima of population transfer across the ion-beam cross-section, as can be seen in Fig.\,\ref{fig:SimPopTransRadial} a). 
Secondly, the width of the second Raman transition is $33\,\%$ larger in frequency space (see Tab.\,\ref{tab:params}), but three times as large in energy space. This is due to the lasers counter-propagating in the first interaction but co-propagating in the second interaction, resulting in differential Doppler factor $\Delta D = \partial\delta/\partial E$ of $2\pi\cdot18.0\,$MHz/eV (counter-propagating) and $2\pi\cdot7.7\,$MHz/eV (co-propagating) and thus selections in energy space of $15\,$meV and $45\,$meV, respectively. Again, this results in ions that are not pumped out of the ground state in the first interaction and can be transferred in the second interaction, reducing the distinctness of the Lamb dip, as shown in Fig.\,\ref{fig:SimPopTransRadial} b). 
This softening of the Lamb dip can also be seen in the simulated spectra in Fig.\,\ref{fig:Sim_Spectra}. Like in Fig.\,\ref{fig:Sim_filter_eff}, simulations were performed for $\Delta\delta\overset{\wedge}{=}\Delta U \in \{0, 0.01\,\mathrm{V}, 0.1\,\mathrm{V}\}$. Even when neglecting fluctuations in the interaction potential and the finite laser line-width ($\Delta\delta = 0$), the relative strength of the Lamb dip is limited to about $20\%$ of the resonance signal. Increasing $\Delta\delta$ results in a broadening of the Lamb dip and consequently, a reduction of the dip strength since the population transfer is distributed over a wider range of ion energies. To distinguish whether $\Delta\delta$ results from the finite laser line-width (in the ion rest frame) or fluctuations in the interaction potential, a single Raman transition does not suffice. Different transitions with corresponding laser configurations are needed to investigate this. The differential Doppler factor $\Delta D$, and thus, the conversion factor between the line-width in frequency- and energy space depends on both the laser frequencies and their direction. Thus, the broadening induced by the variations in the interaction potential will be constant in energy space but scale with $\Delta D$ in frequency space, and vice versa for the broadening induced by the laser line-width.


\section{\label{Results}Experimental Results}

\begin{figure}
    \centering
    \includegraphics[width=0.75\linewidth]{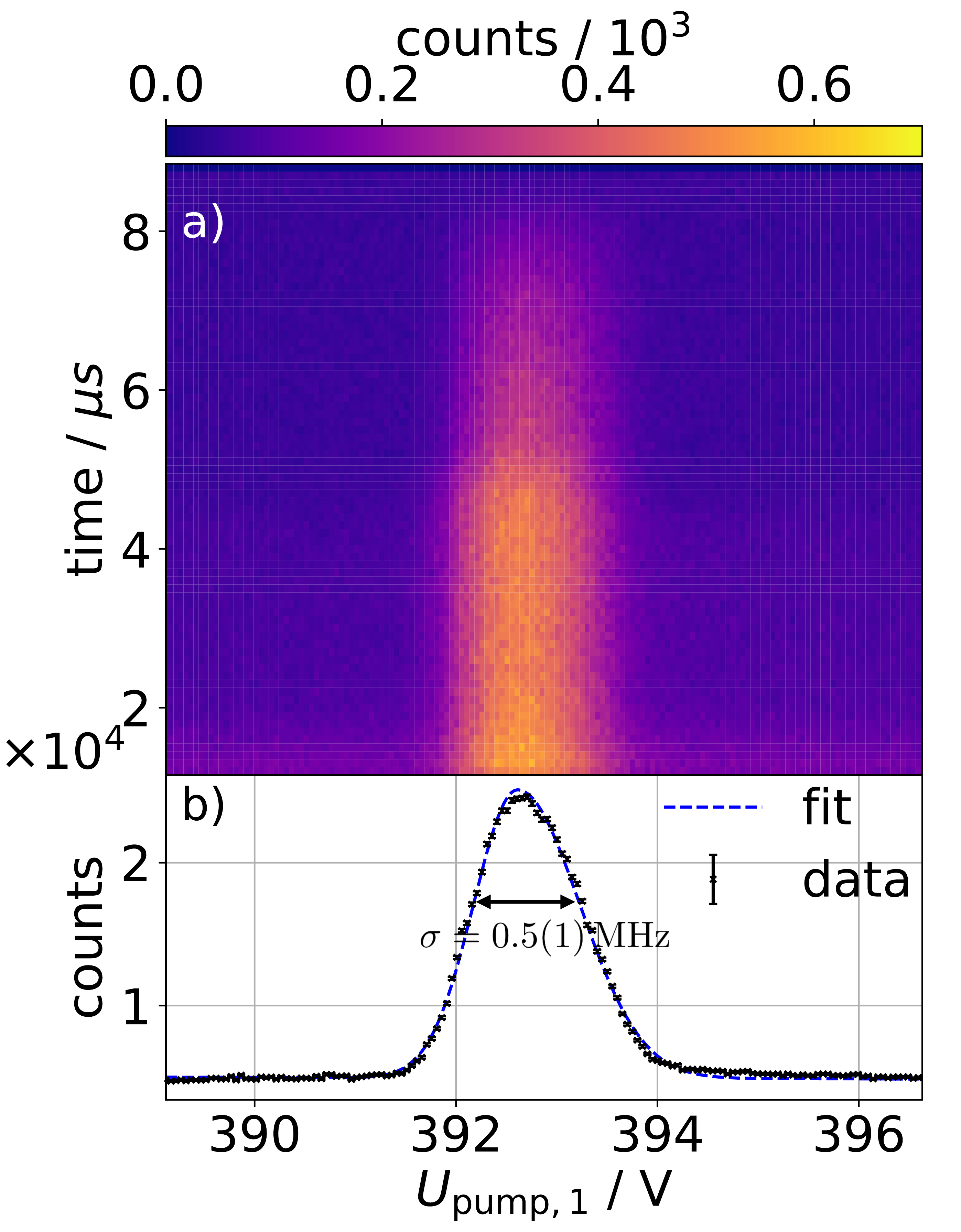}
    \caption{Time-resolved spectrum of the $\mathrm{S}_{1/2}\rightarrow\mathrm{D}_{3/2}$ Raman transition. a): Heat map of the number of measured counts over voltage $U_\mathrm{pump,1}$ applied to the interaction region and time $t$ .The AOM for the 408-nm laser is switched off at $t=0$. At first, ions that interacted with both lasers for the entire $5.8\mu$s required to traverse the interaction region are probed. After $\sim 5\mu$s the interaction time of the ions arriving in the FDR starts decreasing since they did not fully pass the interaction region when the AOM was turned off. b): Projection of the fluorescence signal on the $U_\mathrm{pump,1}$ axis with fit of an asymmetric Gaussian function.}
    \label{fig:TRS_Spec}
\end{figure}

In the following section, the collinear Raman spectroscopy measurements performed at COALA will be presented. First, Rabi oscillations were investigated by performing measurements of the $\mathrm{S}_{1/2}\rightarrow\mathrm{D}_{3/2}$ and $\mathrm{S}_{1/2}\rightarrow\mathrm{D}_{5/2}$ Raman transitions at different experimental parameters. In a second step, the Doppler-free measurement scheme proposed in Sec.\,\ref{sec:MScheme} was implemented to test the feasibility of using Raman transitions as a velocity filter, as proposed by Neumann \textit{et al.}\, \cite{neumann2020raman}.
\\
For both measurements, typical laser powers of $2-3\,$mW, $250-300\,$mW, and $15-20\,$mW were used for the 408-nm, the 1004-nm, and the 1033-nm lasers, respectively. Using a Thorlabs BC106-VIS beam profiler, the beam waists of the lasers were determined to be $w_{408}=1.8\,(1)\,$mm,  $w_{1004}=3.3\,(3)\,$mm, and $w_{1033}=2.6\,(2)\,$mm in the first and $w_{408}=1.7\,(1)\,$mm,  $w_{1004}=3.0\,(2)\,$mm, and $w_{1033}=2.7\,(2)\,$mm in the second interaction region.
\\
An example spectrum of the $\mathrm{S}_{1/2}\rightarrow\mathrm{D}_{3/2}$ resonance at a detuning of $1\,$GHz is shown in Fig.\,\ref{fig:TRS_Spec}. 
The linewidth induced by the energy width of the ion beam still exceeds the intrinsic width of the Raman transition by two orders of magnitude. Thus, the Lorentzian contribution of the expected Voigt profile is assumed to be negligible and an asymmetric Gaussian profile was fitted. 
The asymmetry in the energy distribution is induced by the voltage gradient along the crucible of the surface ionization source, resulting in different starting potentials of the ions \cite{konig2020new}.
\\
Based on measurements performed with an infrared camera, an ion-source temperature of $2100-2300\,$K can be estimated, corresponding to an energy width of $0.46\,(2)$V, which is in good agreement with the observed line-width of $0.5\,(1)\,$eV or $3.9\,(7)\,$MHz.

\subsection{\label{RabiOsc}Collapse of Rabi Oscillations}
\begin{figure}
    \centering
    \includegraphics[width=.85\linewidth]{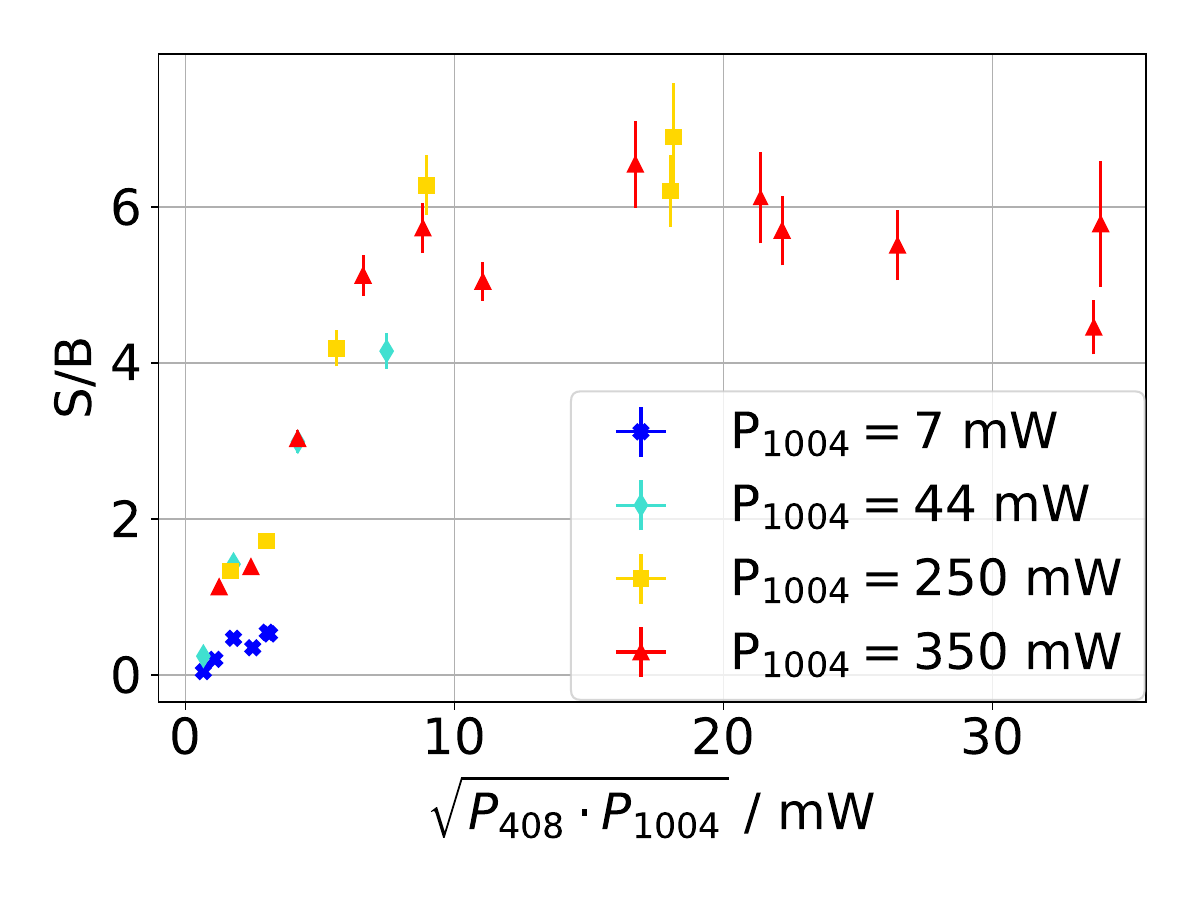}
    \caption{Power dependence of the  signal-to-background (S/B) ratio  of the $\mathrm{S}_{1/2}\rightarrow\mathrm{D}_{3/2}$ resonance, measured at a detuning of $\Delta = 300\,$MHz using the single Raman measurement scheme shown in Fig.\,\ref{fig:Beamline+MSchemes} b). $P_{1004(408)}$ is the power of the 1004-nm (408-nm) laser, respectively}
    \label{fig:CRO_P}
\end{figure}

The simulations showed that the spatial intensity distributions of the ion and laser beams lead to a collapse of the Rabi oscillation, the phase of which scales with $\sqrt{P_\mathrm{L1}P_\mathrm{L2}}\,t/\Delta$.
As the PMT signal scales with the population of the metastable state, this effect can be investigated by performing measurements at different laser powers, interaction times, and detunings. Measurements were performed in both the $\mathrm{S}_{1/2}\rightarrow\mathrm{D}_{3/2}$ and the $\mathrm{S}_{1/2}\rightarrow\mathrm{D}_{5/2}$ Raman resonance, yielding qualitatively identical results. Therefore, only the results in the $\mathrm{S}_{1/2}\rightarrow\mathrm{D}_{3/2}$ transition are discussed.
\\
Fig.\,\ref{fig:CRO_P} shows the signal-to-background (S/B) ratio of the $\mathrm{S}_{1/2}\rightarrow\mathrm{D}_{3/2}$ resonance at a detuning of $\Delta = 300\,$MHz for different fixed powers $P_\mathrm{1004}=7$, $44$, $250$, and $350\,$mW of the 1004-nm laser and for a varying power $P_\mathrm{408}$ of the 408-nm laser from $10\,\mu$W to $3\,$mW as a function of $\sqrt{P_\mathrm{1004}P_\mathrm{408}}$. As the probing efficiency of the metastable state via the $\mathrm{D}_{3/2}\rightarrow\mathrm{P}_{3/2}$ dipole transition depends on the intensity of the 1004-nm laser, potential impacts on the S/B were considered. For $P_\mathrm{1004}\geq44\,$mW, the S/B ratio at fixed $\sqrt{P_\mathrm{1004}P_\mathrm{408}}$ is independent of $P_\mathrm{1004}$, as the probing transition is saturated. Thus, all variations in the S/B ratio for $P_\mathrm{1004}\geq44\,$mW can be ascribed to the efficiency of the population transfer via the Raman transition. No oscillation in the S/B ratio can be seen, only an increase up to a value of $\sqrt{P_\mathrm{1004}P_\mathrm{408}}\approx15\,$mW, above which the S/B ratio saturates.

\begin{figure}
    \centering
    \includegraphics[width=1\linewidth]{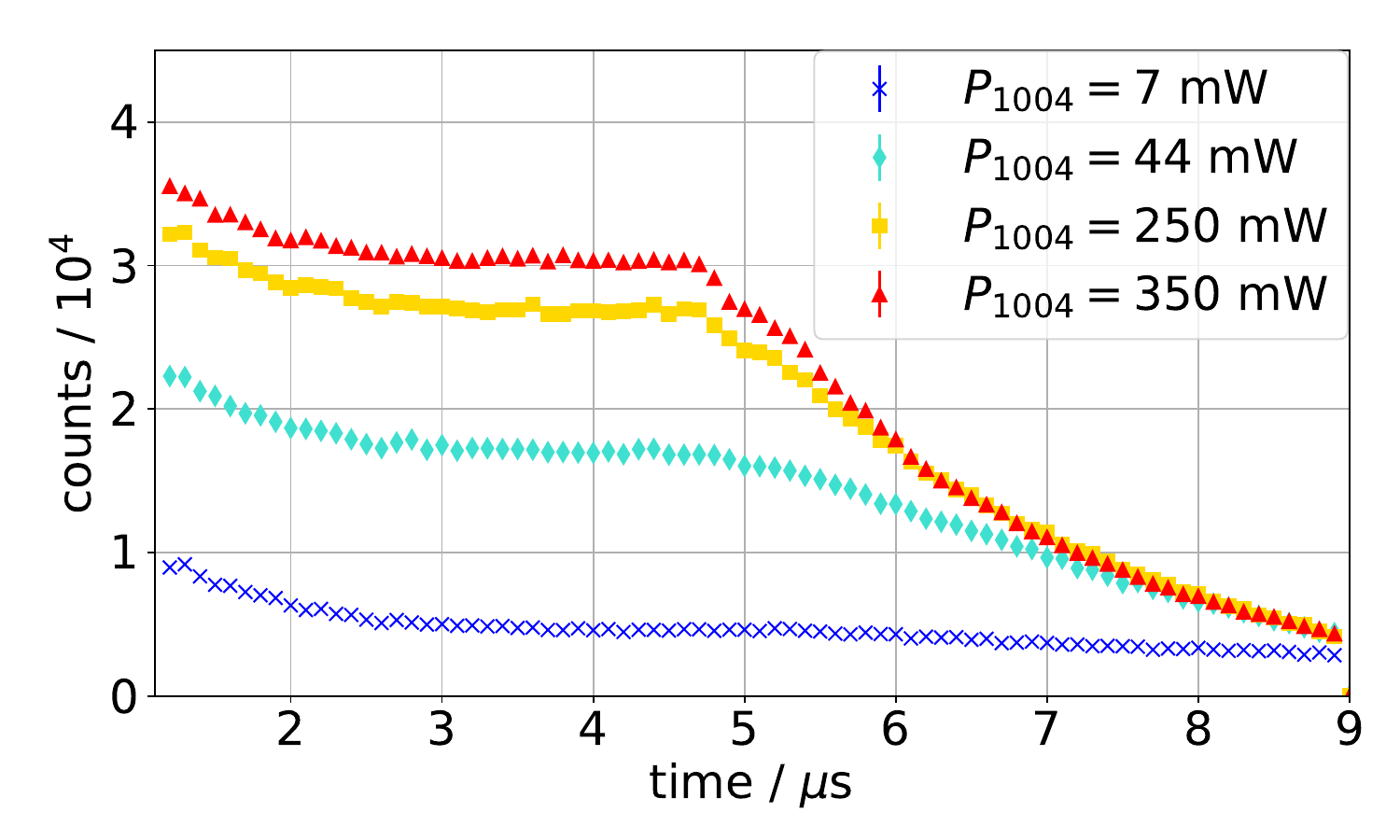}
    \caption{ Evolution of the $\mathrm{S}_{1/2}\rightarrow\mathrm{D}_{3/2}$ resonance signal with time for different powers $P_{1004}$ of the 1004-nm laser and a 408-nm laser power of $P_{408}=1.3\,(1)\,$mW. Shown are the PMT counts over time, starting at the moment the 408-nm laser was turned off with the AOM. The decrease in signal during the first $2\,\mu$s is caused by the finite fall time of the AOM. The phase of the Raman transition of the ions arriving after $4.8\,\mu$s decreases, as the ions have not experienced the full interaction time when the AOM is turned off.}
    \label{fig:CRO_t}
\end{figure}

The time dependence of the resonance signal is plotted in Fig.\,\ref{fig:CRO_t} for $P_\mathrm{1004}=7$, $44$, $250$, and $350\,$mW and $P_\mathrm{408}=1.3\,(1)\,$mW. The 408-nm laser is switched off at $t=0$. As the time-of-flight from the end of the first interaction region to the FDR is approximately $4.8\,\mu$s, ions probed in the FDR in the first $4.8\,\mu$s interacted with both lasers over the entire length of the pumping tube, resulting in a constant signal. 
The increased signal during the first $2\,\mu$s is ascribed to scattered photons of the blue beam during the finite fall time of the AOM transmission. After $4.8\,\mu$s, the probed ions experienced a reduced interaction time, as the 408-nm laser was switched off when they had only partially crossed the interaction region. This results in a linear decrease of the phase of the two-photon Rabi oscillation. However, no oscillation in the PMT counts can be seen for a declining interaction time, even for a theoretical $8\pi$-pulse at $P_\mathrm{1004}=350\,$mW.
\\
Similar additional measurements were performed at different detunings between $\Delta = 50\,$MHz and $\Delta = 3.5\,$GHz. Again, no oscillation in the S/B ratio was seen.
These observations are consistent with the collapse of the Rabi oscillation due to the spatial intensity distribution of the ion beam and are congruous with the corresponding simulations in Sec.\,\ref{sec:Simulations} and the simulations performed by Neumann \textit{et al.} \cite{neumann2020raman}.

\subsection{\label{ssec:VelFilter}First Realization of a Raman Velocity Filter}

\begin{figure*}
    \centering
    \includegraphics[width=.90\linewidth]{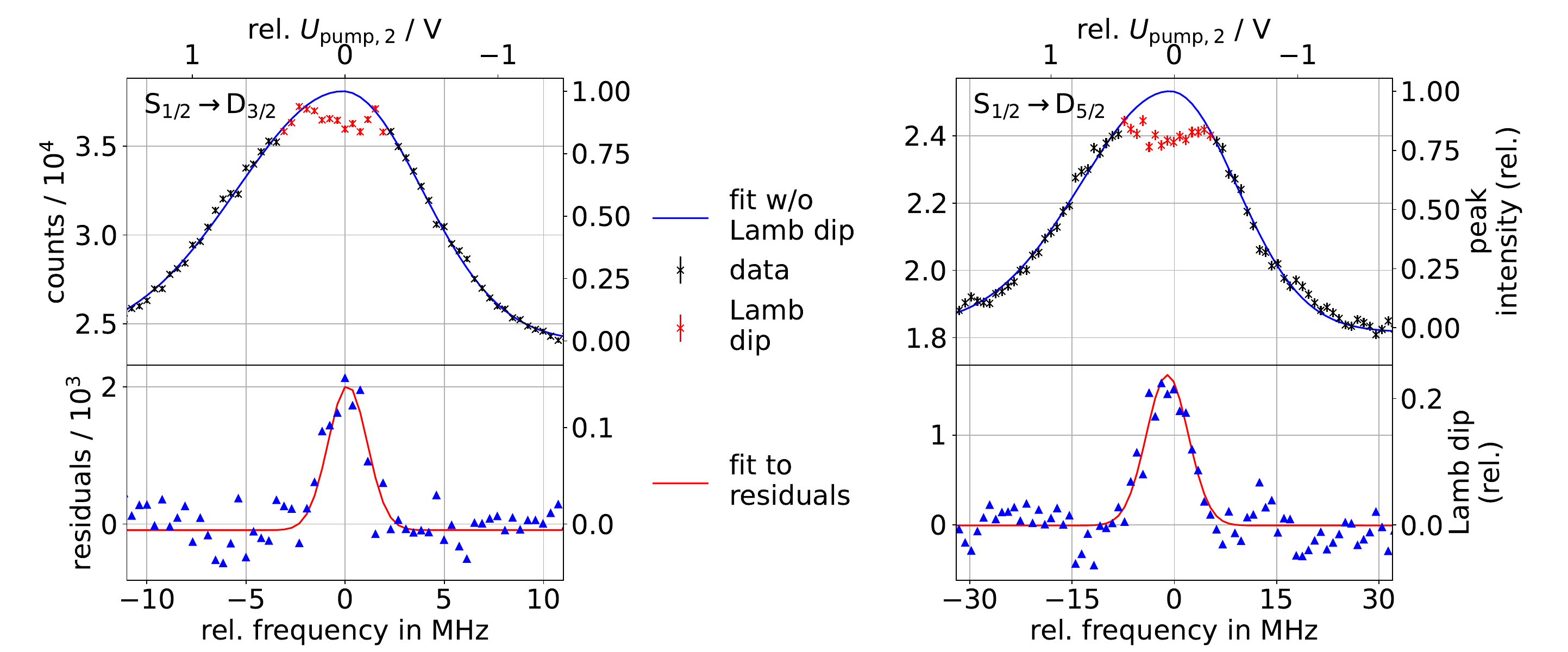}
    \caption{Lamb dip in the $\mathrm{S}_{1/2}\rightarrow\mathrm{D}_{3/2}$ (left) and in the $\mathrm{S}_{1/2}\rightarrow\mathrm{D}_{5/2}$ (right) Raman transition. A first  $\mathrm{S}_{1/2}\rightarrow\mathrm{D}_{5/2}$ transition is driven at a fixed voltage in first interaction region. The voltage of the second interaction region, see Fig.\,\ref{fig:Beamline+MSchemes} a), is scanned across the respective transition as described in Sec.\,\ref{sec:MScheme}. On the lower $x$-axis, the transition frequency is given relative to $436\,373\,693\,$MHz and $444\,779\,044\,$MHz, and on the upper $x$-axis $U_\mathrm{pump,2}$ is given relative to $265.85\,$V and $390.58$V, respectively. The left $y$-axis shows the measured PMT-counts and the right $y$-axis the relative signal strength, analogously to Fig.\,\ref{fig:Sim_Spectra}.
    The Lamb dip becomes visible as a peak in the residuals of the fit to the resonance excluding  data points lying on the Lamb dip, which are indicated in red, shown in the lower part of the figure. 
    }
    \label{fig:DoubleRamanSpectra}
\end{figure*}

To investigate the potential capabilities of a Raman velocity filter, the Lamb dip in the ground-state population, induced by a first $\mathrm{S}_{1/2}\rightarrow\mathrm{D}_{5/2}$ Raman transition, was measured in the resonance signal of a second, sequential  $\mathrm{S}_{1/2}\rightarrow\mathrm{D}_{5/2}$ or  $\mathrm{S}_{1/2}\rightarrow\mathrm{D}_{3/2}$ Raman transition using the Doppler-free scheme described in Sec.\,\ref{sec:MScheme}. Example spectra of both resonances with Lamb dip are shown in Fig.\,\ref{fig:DoubleRamanSpectra}.
The $\mathrm{S}_{1/2}\rightarrow\mathrm{D}_{5/2}$ Raman transition was driven at a detuning of $\Delta=-665\,$MHz and the $\mathrm{S}_{1/2}\rightarrow\mathrm{D}_{3/2}$ at a detuning of $1020\,$MHz.
\\
The line shape of the Lamb dip was obtained by first fitting the resonance signal, excluding the data points lying on the Lamb dip. For this, again, an asymmetric Gaussian was chosen. The parameters of this fit are constrained to be consistent with spectra without Lamb dip. The residuals of this fit then allow us to investigate the Lamb dip, as shown in the lower frame in Fig.\,\ref{fig:DoubleRamanSpectra}. Different line shapes (Gaussian/Lorentzian/Voigt) were fitted to the Lamb dip. 
In the scenario of an optimal velocity filter, which selects only one velocity class in the first interaction, one would expect the line shape of the Lamb dip to be given by the intrinsic line shape of the (second) Raman transition, ergo a Lorentzian. However, based on the vanishing Lorentzian contribution of the Voigt fit, the measured Lamb dip seems to be best described by a Gaussian. Here, further insight can be gained by comparing the width of the Lamb dip in the different Raman transitions:
In spectra of the $\mathrm{S}_{1/2}\rightarrow\mathrm{D}_{5/2}$ transition, driven by the anticollinear 408-nm laser and the collinear \mbox{1033-nm} laser, the width of the Lamb dip is  $3.3\,(2)\,$MHz, and for the $\mathrm{S}_{1/2}\rightarrow\mathrm{D}_{3/2}$ transition, driven by the anticollinear 408-nm laser and the anticollinear 1004-nm laser, $1.0\,(1)\,$MHz. The ratio of those widths approximately corresponds to that of the differential Doppler factors of $7.8\,$MHz/eV and $18.0\,$MHz/eV for these transitions. 
The width of the Lamb dip and the inferred ion energy distribution is $0.15(5)\,$eV. 
and is limited by fluctuations in the interaction potential of $\delta\Delta \leq 0.15\,(5)\,$V. 
This also fits the observed relative strength of $10\%$--$20\%$ of the Lamb dip, consistent with simulations for $\delta \approx 0.1\,$V. The fluctuations could be caused by the noise of the power supply used, as well as by potential gradients and field penetrations at the entrance and exit of the interaction regions.
Additional contributions that were considered are time-of-flight broadening, which is especially relevant in the second, shorter interaction region, as well as the laser line-widths and the relative stability of the laser frequencies. The latter was investigated by measuring the beat signal between the different lasers at $1033\,$nm but could only be constrained to be less than $0.5$\,MHz.
\\

\section{\label{Outlook}Conclusion \& Outlook}

We demonstrated optical Raman transitions in collinear laser spectroscopy on a fast ion beam using two separate lasers with linewidths of a few $100\,$kHz to address dipole-forbidden fine-structure transitions.
Simulations for consecutive Raman transitions in $^{88}$Sr$^+$, including realistic ion energy distributions, correctly predict the width of the observed resonance and Lamb dip, as well as the collapse of two-photon Rabi oscillations. The latter is caused by the spatial intensity distributions of laser and ion beams are taken into account.\\
First measurements using collinear Raman saturation spectroscopy were performed to investigate the feasibility of using Raman transitions as a velocity filter.
Employing the Raman velocity filter allowed us to reduce the effective energy width of ions excited to the $4d$ state to $0.15\,(5)\,$eV, corresponding to $7.5\,(2.5)\,$ppm of the total beam energy. We note that the initial energy width of $0.5\,(1)\,$eV is already small since we used a surface ion source, while it would be considerably larger if a liquid metal ion source, an electron beam ion source or a laser ablation ion source is used, which are all available at COALA \cite{konig2020new,Imgram2023} and have much larger energy spreads. Using ion species from these sources, a Raman velocity filter would provide much larger reduction factors.\\
Investigating Raman transitions in different geometries, allowing to differentiate between limitations in frequency and voltage space, shows that this could be improved by using power supplies with a higher stability and minimizing field penetration into the interaction regions.\\
The application of the Raman transition provides significant improvements compared to an allowed dipole transition for velocity-selective excitation, like the $4s\,^2\mathrm{S}_{1/2}\rightarrow 5p\,^2\mathrm{P}_{1/2,3/2}$ transition in Ca$^+$ used in our previous high-voltage measurements \cite{kramer2018high}. 
In Sr$^+$, for example, the corresponding $5s\,^2\mathrm{S}_{1/2}\rightarrow 5p\,^2\mathrm{P}_{1/2,3/2}$ transition has a linewidth of $22.4\,$MHz and -- at the beam energies used in our measurements presented here -- a differential Doppler factor of this transition is $12.9\,$eV. This determines an acceptance window of about $1.74$\,eV, being more than an order of magnitude larger than the $150\,$meV demonstrated for the Raman filter.

The observed linewidth of $1.0\,(1)$\,MHz for the measured Lamb dip is comparable to the relative energy width of $3.3\,$ppm and Lamb-dip width of $1.6\,$MHz estimated by Neumann \textit{et al.} \cite{neumann2020raman} and confirms the feasibility of using a Raman velocity filter for future high-voltage measurements at COALA. 


\begin{acknowledgments}
We thank the group of Ruben de Groote from KU Leuven for providing the diode laser without that many of the performed measurements would not have been possible.
We acknowledge support from the German Research Foundation (DFG, Deutsche Forschungsgemeinschaft) under project number NO789/4-1 and -- Project-ID 279384907 -- SFB 1245, as well as by the German Federal Ministry of Research, Technology and Space (BMFTR) under contract No. 05P24RD8. J.S. and H.B. acknowledge support from HGS-HIRe.
\end{acknowledgments}

\appendix

\section{\label{secAppendA}Transformation of the three-level Hamiltonian to an effective two-level Hamiltonian}

The Hamiltoninan $\hat{H}_\mathrm{RWA}$ of a three-level system interacting with two lasers as shown in Fig.\,\ref{fig:RamanLvlscheme},
using minimal coupling, and applying the Power-Zienau-Woolley transformation and the rotating wave approximation, can be decomposed into a kinetic, an atomic, and an interaction part, given by

\begin{subequations}
    \begin{align}
        \begin{split}
        \label{eq:H_kin}
            \hat{H}_{\mathrm{kin}} &= \left( \begin{array}{ccc}
                        \frac{\Vec{p}^2}{2M} & 0 & 0 \\
                        0 & \frac{(\Vec{p}+\hbar(\Vec{k}_{\mathrm{L}1}-\Vec{k}_{\mathrm{L}2})^2}{2M} & 0 \\
                        0 & 0 & \frac{(\Vec{p}+\hbar \Vec{k}_{\mathrm{L}1})^2}{2M}
                    \end{array} \right),
        \end{split}\\
        \begin{split}
        \label{eq:H_atm}
            \hat{H}_{\mathrm{atm}} &= \left( \begin{array}{ccc}
                    \hbar \omega_1 & 0 & 0 \\ 
                    0 & \hbar \omega_2 & 0 \\
                    0 & 0 & \hbar \omega_3
                \end{array} \right),
        \end{split}\\
        \begin{split}
        \label{eq:H_int}
            \hat{H}_{\mathrm{int}} &= \left( \begin{matrix}
                    0 & 0 & \frac{\hbar\Omega_{11}}{2} e^{i(\omega_{\mathrm{L}1}t)}\hspace{-3em} \\
                    0 & 0 & \frac{\hbar\Omega_{22}}{2} e^{i(\omega_{\mathrm{L}2}t)}\hspace{-3em} \\
                    \frac{\hbar\Omega_{11}^*}{2} e^{-i(\omega_{\mathrm{L}1}t)}\hspace{-3em} & \frac{\hbar\Omega_{22}^*}{2} e^{-i(\omega_{\mathrm{L}2}t)}\mkern-40mu & 0
                \end{matrix} \right).
        \end{split}
    \end{align}
\end{subequations}

The interaction-frame Hamiltonian in Eq.\,\ref{eq:H_IF} is directly obtained via a unitary transformation  $\ket{\Psi}_\mathrm{IF} = \begin{pmatrix} b_1 & b_2 & b_3 \end{pmatrix}^T = \hat{U}\ket{\Psi}_\mathrm{RWA}$ with $\hat{U}=  e^{i(\hat{H}_{\mathrm{kin}}+\hat{H}_{\mathrm{atm}})t/\hbar}$.\\
Writing out the time-dependent Schrödinger equation yields a system of coupled differential equations,
\begin{subequations}
    \begin{align}
        \begin{split}
        \label{eq:db1/dt}
            \Leftrightarrow i\hbar\frac{\partial}{\partial t}b_1(t) &= \frac{\hbar\Omega_{11}}{2}e^{-i(\Delta t+\phi_1)}b_3(t)
        \end{split}\\
        \begin{split}
        \label{eq:db2/dt}
            i\hbar\frac{\partial}{\partial t}b_2(t) &= \frac{\hbar\Omega_{22}}{2}e^{-i((\Delta+\delta) t+\phi_2)}b_3(t)
        \end{split}\\
        \begin{split}
        \label{eq:db3/dt}
            i\hbar\frac{\partial}{\partial t}b_3(t) &= \frac{\hbar\Omega_{11}^*}{2}e^{i(\Delta t+\phi_1)}b_1(t)\\ 
            &+ \frac{\hbar\Omega_{22}^*}{2}e^{i((\Delta+\delta) t+\phi_2)}b_2(t),
        \end{split}
    \end{align}
\end{subequations}
where $\phi_{1,2}$ are phases of the laser fields that have been added for completeness. Assuming $\Delta \gg \delta$, $\frac{\partial}{\partial t}b_1(t)$, $\frac{\partial}{\partial t}b_2(t) \sim \Omega_{ii}$, Eq.\,\eqref{eq:db3/dt} can be integrated, assuming $b_1(t)$, $b_2(t)$ to oscillate slower than $e^{i\Delta t}$ and thus constants in the integral, in order to obtain $b_3(t)$. 
Inserting $b_3(t)$ into Eq.\,\eqref{eq:db1/dt} and \eqref{eq:db2/dt} yields\\

\begin{subequations}
    \begin{align}
        \begin{split}
            \label{eq:b1approx}
            i\hbar\frac{\partial}{\partial t}b_1(t) &= -\frac{\hbar|\Omega_{11}|^2}{4\Delta}b_1(t) \\
            &-\frac{\hbar\Omega_{11}\Omega_{22}^*}{4(\Delta+\delta)}e^{i(\delta t - \phi_{L})}b_2(t),
        \end{split}\\
        \begin{split}
            \label{eq:b2approx}
            i\hbar\frac{\partial}{\partial t}b_2(t) &= -\frac{\hbar\Omega_{22}\Omega_{11}^*}{4\Delta}e^{-i(\delta t- \phi_{L})}b_1(t) \\
            &-\frac{\hbar|\Omega_{22}|^2}{4(\Delta+\delta)}b_2(t),
        \end{split}
    \end{align}
\end{subequations}
where $\phi_L=\phi_1 - \phi_2$ is the relative phase of the two lasers. Since the detuning from both single-photon resonances is large, the population of $\ket{3}$ is negligible and Eqs.\,\eqref{eq:b1approx} and \eqref{eq:b2approx} can be used to obtain an effective two-level Hamiltonian, given by

\begin{equation}
    \label{eq:H2.0}
   \hat{H}_\mathrm{IF} = -\hbar \begin{pmatrix}
       \Omega_1^\mathrm{AC} & \frac{\Omega_{\mathrm{R}2}}{2}e^{i(\delta t - \phi_{L})}\\
        \frac{\Omega_{\mathrm{R}1}^*}{2}e^{-i(\delta t - \phi_{L})} & \Omega_2^\mathrm{AC}
    \end{pmatrix}
\end{equation}
with 
\begin{eqnarray}
    \Omega_{\mathrm{R}n} = \Omega_{11}\Omega_{22}^*/2\Delta_{nn} \stackrel{\Delta \gg \delta}{\approx} \Omega_{11}\Omega_{22}^*/2\Delta =: \Omega_R
\end{eqnarray}
The diagonal terms are the first order AC-Stark shifts $\Omega_n^{AC}$ of the state $\ket{n}$
\begin{equation}
    \label{eq:Omega_AC}
    \Omega_n^{\mathrm{AC}} = \sum_{j=1,2}\frac{|\Omega_{nj}|^2}{4\Delta_{nj}}, \quad n \in \{1,2\}, 
\end{equation}
where the second term results from the so far neglected far off-resonant couplings $\Omega_{nj}$ with $n\neq j$. Note that all terms scale with $\frac{1}{\Delta_{nj}}$ and that thus all contributions to both the AC-Stark shift and the time evolution of the system resulting from those couplings can be omitted.\\
Applying two more unitary transformations 
\begin{equation}
    \label{eq:U2.1}
    \hat{U}=\begin{pmatrix}
   e^{-i\Omega_1^\mathrm{AC}} & 0\\
   0 & e^{-i\Omega_2^\mathrm{AC}}
    \end{pmatrix}
\end{equation}
and 
\begin{equation}
    \label{eq:U2.2}
    \hat{U}=\begin{pmatrix}
   e^{-i\frac{(\delta-\delta_\mathrm{AC})}{2}t} & 0\\
   0 & e^{i\frac{(\delta-\delta_\mathrm{AC})}{2}t}
    \end{pmatrix}\mathrm{,}
\end{equation}
with $\delta_\mathrm{AC} = \Omega_1^\mathrm{AC}-\Omega_2^\mathrm{AC}$, to first eliminate diagonal elements and then eliminate time dependencies, yields a time-independent two-level Hamiltonian 
\begin{equation}
 \label{eq:H2.3}
   \hat{H}_\mathrm{2LVL} = -\hbar \begin{pmatrix}
       (\delta-\delta_\mathrm{AC})/2 & \frac{\Omega_{\mathrm{R}}}{2}e^{-i\phi_{L}}\\
        \frac{\Omega_{\mathrm{R}}^*}{2}e^{i\phi_{L}} & -(\delta-\delta_\mathrm{AC})/2 
    \end{pmatrix}\mathrm{.}
\end{equation}
This Hamiltonian has the form of an ideal two-level system interacting with a single laser. Tthe Rabi frequency is now given by the two-photon Rabi frequency $\Omega_\mathrm{R}$ and the detuning is given by the two-photon detuning, corrected for the AC-Stark shift. This also implies that this effective two-level system can be described on the Bloch sphere or by using the dressed-state approach and that the dynamics of the system is described by the Rabi oscillation in Eq.\,\ref{eq:P2(t)}.

\nocite{*}

\bibliography{ARamanVelocityFilter}

\end{document}